\documentclass[twocolumn,preprintnumbers,prl,aps]{revtex4-1}
\usepackage{graphicx,amsmath,bbm,color,amssymb,amscd}
\usepackage[breaklinks=true]{hyperref}
\graphicspath{ {fig/} }

\allowdisplaybreaks[4]

\definecolor{darkblue}{rgb}{0.0,0.0,0.5}

\newcommand{\graph}[2][0.3]{\vcenter{\hbox{\includegraphics[scale=#1]{#2}}}}
\newcommand{\dimgraph}[2]{\overset{(#2-2\epsilon)}{\graph[0.2]{#1}}}
\newcommand{\figgraph}[3]{\overset{(#3-2\epsilon)}{\graph[#1]{#2}}}

\begin{document}

\preprint{MSUHEP-18-009}

\title{\boldmath
Constructing multi-loop scattering amplitudes with manifest singularity structure
\unboldmath}

\author{Robert M. Schabinger} 

\affiliation{Department of Physics and Astronomy, Michigan State University, East Lansing, Michigan 48824, USA}

\begin{abstract}
\noindent
The infrared exponentiation properties of dimensionally-regularized multi-loop scattering amplitudes are typically hidden at the level of the integrand, materializing only after integral evaluation. We address this long-standing problem by introducing an appropriate integral basis which is simultaneously finite and uniform weight. As an example, we cast the integrand for the QCD corrections to the two-loop massless quark electromagnetic form factor into a form where it is manifest that the $\epsilon^{-4}$ and $\epsilon^{-3}$ pole terms arise from the $\epsilon$ expansion of the square of our one-loop master integral.
\end{abstract}

\maketitle

\section{introduction}
Our understanding of the infrared structure of dimensionally-regularized gauge theory scattering amplitudes involving massless quanta is still not complete, despite decades of research \cite{Brandt:1982gz,Gatheral:1983cz,Frenkel:1984pz,Korchemsky:1985xj,Sterman:1986aj,Korchemsky:1987wg,Collins:1989gx,Korchemsky:1988hd,Korchemsky:1988si,Magnea:1990zb,Korchemskaya:1994qp,Catani:1996jh,Contopanagos:1996nh,Kidonakis:1997gm,Kidonakis:1998bk,Catani:1998bh,Kidonakis:1998nf,Sterman:2002qn,Dokshitzer:2005ig,Aybat:2006wq,Aybat:2006mz,Mitov:2006xs,Dixon:2008gr,Becher:2009cu,Gardi:2009qi,Becher:2009qa,Becher:2009kw,Dixon:2009ur,Gardi:2010rn,Bret:2011xm,DelDuca:2011ae,Ahrens:2012qz,Dukes:2013wa,Gardi:2013ita,Caron-Huot:2013fea,Falcioni:2014pka,Boels:2017skl,Almelid:2015jia,Henn:2016jdu,Caron-Huot:2017zfo,Almelid:2017qju,Moch:2017uml,Grozin:2017css,Caron-Huot:2017zfo,Boels:2017ftb}. One curious feature of multi-loop hard scattering amplitudes is that, at least in all realistic models, they do not exhibit any physically-meaningful pole structure when written in integral form with respect to any known Feynman integral basis. This is surprising because one would naively think that the higher $\epsilon$ pole terms should have their origin in factorizable topologies; Catani \cite{Catani:1998bh} explained long ago how to predict the $\epsilon^{-4}$ and $\epsilon^{-3}$ pole terms of quite general on-shell two-loop scattering amplitudes from the corresponding one-loop results and appropriate universal one-loop quantities. We show in this paper that it is possible to make manifest at least this aspect of gauge theory infrared exponentiation at the integrand level, simply by making a more appropriate choice of integral basis.

The key idea is to combine two extant Feynman integral basis paradigms to construct a new integral basis which is simultaneously finite \cite{vonManteuffel:2014qoa,vonManteuffel:2015gxa} and uniform weight \cite{Kotikov:2010gf,Henn:2013pwa}. Note that extensions of the original uniform weight basis construction intended to be applicable to general scattering processes ({\it i.e.} beyond the realm of multiple polylogarithms) have been proposed in the literature \cite{Bonciani:2016qxi,vonManteuffel:2017hms,Ablinger:2017bjx,Remiddi:2017har,Bluemlein:2017ibj,Bourjaily:2017bsb,Hidding:2017jkk,Broedel:2017kkb,Broedel:2017siw,Adams:2018yfj,Broedel:2018iwv,Lee:2018jsw,Broedel:2018qkq}. In this paper, we study processes where the integral basis is polylogarithmic in nature, though it seems likely to us that our construction will ultimately generalize to processes with richer analytic structures, such as the production of a top-antitop pair in proton-proton collisions (see {\it e.g.} \cite{Adams:2018bsn,Adams:2018kez}). A fully general construction would be of independent interest as a bookkeeping device for scattering amplitude computations because it would deliver Feynman integrals with near-optimal properties, perfect for both analytical and numerical explorations (see {\it e.g.} \cite{vonManteuffel:2017myy} to gain perspective on this point). We will call integrals which are both finite and uniform weight {\it uniformly finite} below.

In the next section, we motivate our construction by studying selected master integrals for the unrenormalized one-loop quark form factor and the one- and two-loop four-point gluon scattering amplitudes of massless QCD. We then show how not all uniformly-finite integral bases are equally interesting and explain how to construct a good uniformly-finite basis for the unrenormalized two-loop quark form factor of massless QCD. Finally, we discuss our result, highlighting the ways in which it is superior to previously-known representations. We conclude with an outlook and some remarks about the potential of our new integral basis paradigm.

\section{motivation}
In this section, we suggest how one could first discover non-trivial uniformly-finite integrals. First, note that it {\it is} trivial to build a uniformly-finite Feynman integral out of any finite integral which evaluates to a single power product of Gamma functions \cite{Lee:2009dh,Henn:2013fah}. Consider the one-loop form factor master integral from \cite{vonManteuffel:2015gxa}:
\begin{align}\label{eq:ff1L}
	\dimgraph{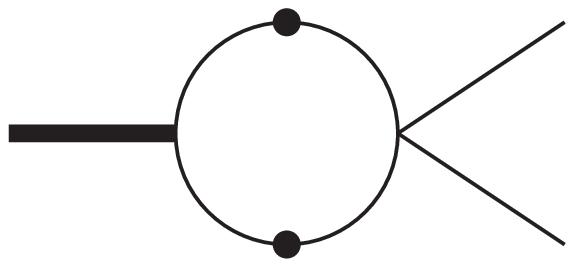}{6} 
	&= 
	\left. \frac{e^{\epsilon \gamma_E}}{i \pi^{3-\epsilon}}\int\!\!\frac{\mathrm{d}^{6-2\epsilon}k}{((p_1+k)^2)^2((p_2-k)^2)^2}~\right|_{q^2 = -1}
	\nonumber \\
	&= \frac{e^{\epsilon \gamma_E}\Gamma^2(1-\epsilon)\Gamma(1+\epsilon)}{\Gamma(2-2\epsilon)}
	\nonumber \\
	&= 1 + 2\epsilon + \left(4 - \frac{1}{2}\zeta_2 \right) \epsilon^2 + \mathcal{O}\left(\epsilon^3\right).
\end{align}
In the above, we have set the virtuality $q^2 = (p_1 + p_2)^2$ to $-1$. To construct our preferred uniformly-finite master integral from Eq. (\ref{eq:ff1L}), we simply dress it with a factor of $1 - 2 \epsilon$. Let us remind the reader that it is straightforward to identify finite Feynman integrals belonging to a given integral topology \cite{vonManteuffel:2014qoa} and that there exists a {\tt Reduze 2} \cite{vonManteuffel:2012np,Studerus:2009ye,Bauer:2000cp,fermat} job for exactly this purpose. Depending on the complexity of the problem, it may be non-trivial to carry out the integration by parts reductions required to map conventional integrals onto dimensionally-shifted ones. While this is an important technical problem in its own right, the examples discussed in this work are simple enough that any publicly available reduction code under active development should be able to solve all of the relevant reduction problems in minutes.

The one-loop fully-massless box integral provides a more non-trivial example because its all-orders-in-$\epsilon$ evaluation involves a sum of ${}_2F_1$ functions \cite{vanNeerven:1985xr}. Using a Tarasov dimension shift \cite{Tarasov:1996br} together with integration by parts \cite{Tkachov:1981wb,Chetyrkin:1981qh}, one can derive the identity
\begin{align}\label{eq:oneloopbox}
&\dimgraph{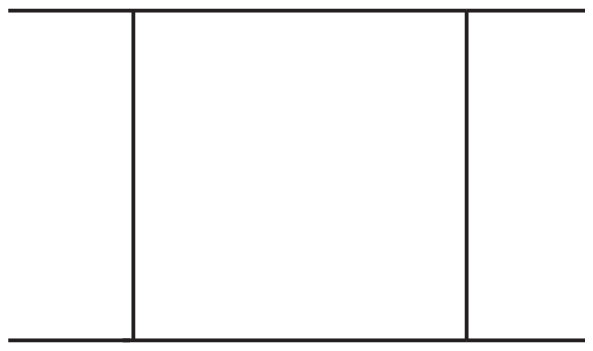}{4} 
= 
\frac{2}{s\,t}\Bigg((s+t)(1-2\epsilon)\dimgraph{oneloopbox}{6}
\\
&- \frac{1}{\epsilon^2}\Bigg(s (1-2\epsilon) \dimgraph{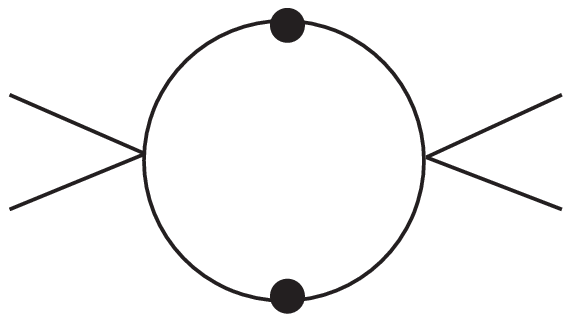}{6}(s) + t (1-2\epsilon) \dimgraph{bubbleforbox}{6}(t)\Bigg)\Bigg).\nonumber
\end{align}
From Eq. (\ref{eq:ff1L}) and the discussion of $6 - 2\epsilon$ boxes in \cite{Bern:1993kr}, we know that the dressed bubble integrals in (\ref{eq:oneloopbox}) are uniformly finite and that the  box integral on the right-hand side is finite. Now, due to the fact that the $4-2\epsilon$ massless box is well-known to be uniform weight with leading singularity $s\,t$ \cite{Bern:1994zx}, we can immediately conclude from Eq. (\ref{eq:oneloopbox}) that the $6 - 2\epsilon$ massless box dressed with $(s + t)(1 - 2\epsilon)$ is a uniformly-finite integral. 

The form of the result for the uniformly-finite massless box is incredibly simple and we arrived at the result without studying any explicit $\epsilon$ expansions. Instead, we were able to leverage our knowledge of the usual uniform weight integral in $4 - 2 \epsilon$. As one might expect, it is somewhat harder to produce satisfactory, non-factorizable, uniformly-finite integral candidates at the multi-loop level. For illustrative purposes, let us survey the four-point integral topologies at one order higher.

Some of the well-known two-loop master integral topologies, such as the one-loop box with a massless bubble insertion, admit simple uniformly-finite integrals no more complicated than what we have seen so far:
\begin{align}
(s + t)(1-2\epsilon)(1-3\epsilon)\figgraph{.2}{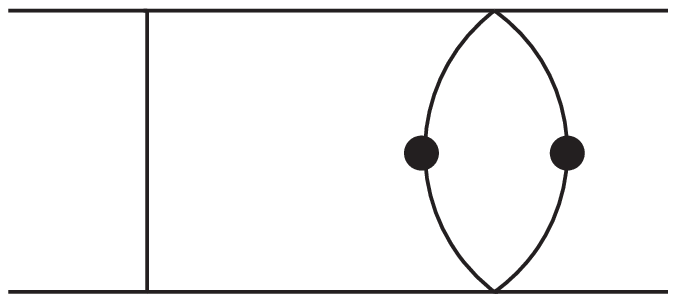}{6}.\nonumber
\end{align}
On the other hand, some topologies, like the two-loop diamond, allow for more complicated uniform-weight linear combinations of finite Feynman integrals:
\begin{align}
(s + t)(1-3\epsilon)(2-3\epsilon)\left(\figgraph{.21}{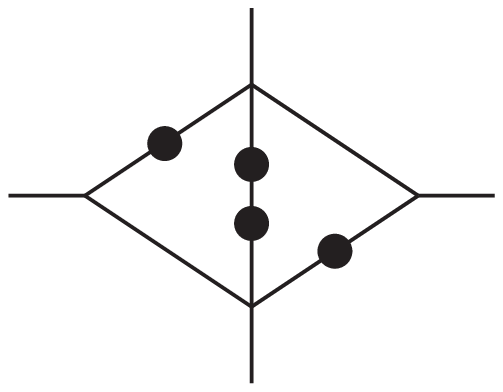}{8}+\figgraph{.21}{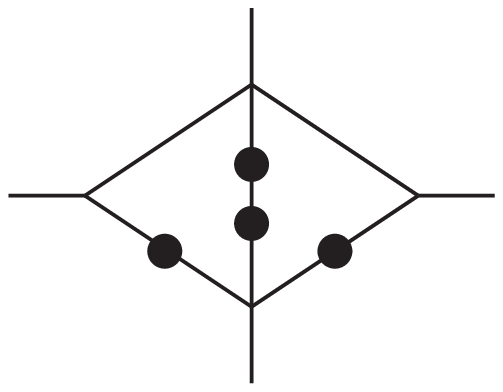}{8}+\figgraph{.21}{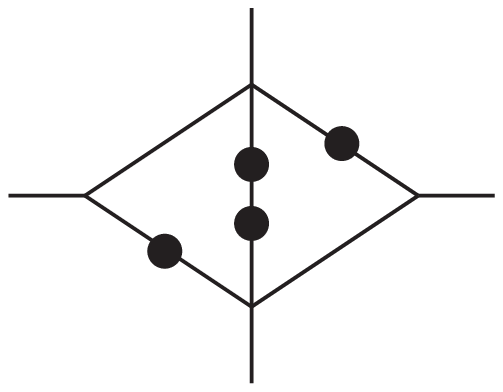}{8}\right).\nonumber
\end{align}
As should be clear from the above examples, the construction of uniformly-finite master integrals at the multi-loop level is non-trivial and must be discussed at length in what follows.
\section{uniformly-finite master integrals for the two-loop quark form factor}
Our goal in this section is to produce an explicit uniformly-finite integral basis which makes manifest the infrared exponentiation properties of the two-loop quark form factor of massless QCD. To that end, let us first review the prediction of the evolution equation satisfied by the bare quark form factor \cite{Magnea:1990zb,Moch:2005id}. If we write the one-loop quark form factor as
\begin{equation}
\mathcal{F}_1^q(\epsilon) = -\frac{\Gamma_1}{2 \epsilon^2} - \frac{G_1(\epsilon)}{2 \epsilon},
\end{equation}
then one finds
\begin{align}
\label{eq:ff2lpred}
&\mathcal{F}_2^q(\epsilon) = \frac{\Gamma_1^2}{8\epsilon^4}+\frac{1}{\epsilon^3}  \bigg(\frac{\Gamma_1 G_1(0)}{4}  - \frac{\beta_0\,\Gamma_1}{8}\bigg)
\\& 
+\frac{1}{\epsilon^2}\bigg(-\frac{\Gamma_2}{8} + \frac{\Gamma_1 G_1^\prime(0) }{4} + \frac{G_1^2(0)}{8} - \frac{\beta_0\, G_1(0)}{4}\bigg) + \cdots,\nonumber
\end{align}
at two-loop order, where
\begin{align}
\label{eq:cusps}
&\Gamma_1 = 4 C_F \quad \Gamma_2 = -\frac{40}{9} C_F N_f + \Big(\frac{268}{9}-8\zeta_2\Big) C_A C_F,
\\& G_1(0) = 6 C_F \qquad G^\prime_1(0) = \Big(16 - 2 \zeta_2\Big) C_F,
\\&{\rm and}~~\beta_0 = \frac{11}{3} C_A - \frac{2}{3} N_f.
\end{align}
In Eq. (\ref{eq:ff2lpred}), it is manifest that the $\epsilon^{-4}$ and $\epsilon^{-3}$ poles are completely fixed by one-loop quantities. Furthermore, from Eq. (\ref{eq:cusps}), we see that the two-loop cusp anomalous dimension, $\Gamma_2$, has no term proportional to the color structure $C_F^2$. Therefore, Eq. (\ref{eq:ff2lpred}) also implies that one-loop quantities dictate the form of the $\epsilon^{-2}$ pole of the $C_F^2$ color structure of $\mathcal{F}_2^q(\epsilon)$. We will see in this section that an integral basis exists where it is manifest that the square of the one-loop master integral from Eq. (\ref{eq:ff1L}) generates all of the structure highlighted above.

Now, as a first step towards this goal, we might try to build uniformly-finite integrals in the two-loop one-external-mass sunrise and the two-loop one-external-mass bubble-triangle  integral topologies. Unfortunately, upon inspecting hundreds of possible finite integral candidates at leading order in $\epsilon$, it becomes clear that both of these topologies admit only trivial uniformly-finite integral candidates which invariably contribute to the $\epsilon^{-4}$ and $\epsilon^{-3}$ pole terms of the unrenormalized two-loop form factor. This is due to the fact that all integral candidates produced by {\tt Reduze 2} may be expressed as single power products of Gamma functions. For example, we can arbitrarily choose the finite basis integrals employed in \cite{vonManteuffel:2015gxa} and write
\begin{widetext}
\begin{align}
\label{eq:badintegrand}
&\mathcal{F}_2^q(\epsilon) = C_F^2 \left(\frac{1}{\epsilon^4} \left[f_1 \left\{ (1-2 \epsilon)^2\figgraph{.15}{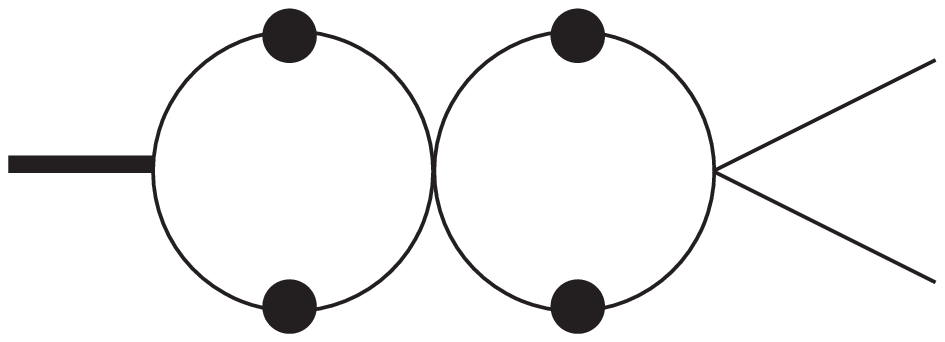}{6}\right\} + f_2 \left\{-8 (1 - 3 \epsilon) (2 - 3 \epsilon)\figgraph{.15}{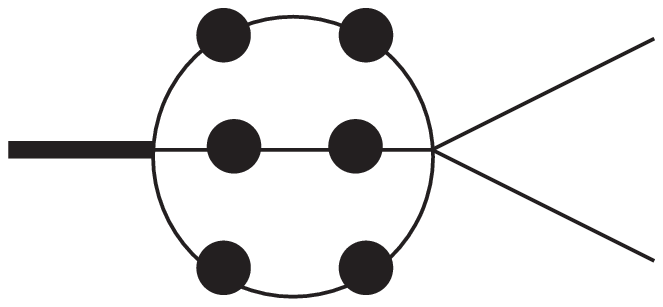}{8}\right\}\right] 
\right. \nonumber \\
& \left. + \frac{f_3}{\epsilon^3} \left\{\frac{32(1-2 \epsilon)(3-2 \epsilon)(1-3 \epsilon)(2-3 \epsilon)}{(1- \epsilon) (1+2 \epsilon)}\figgraph{.15}{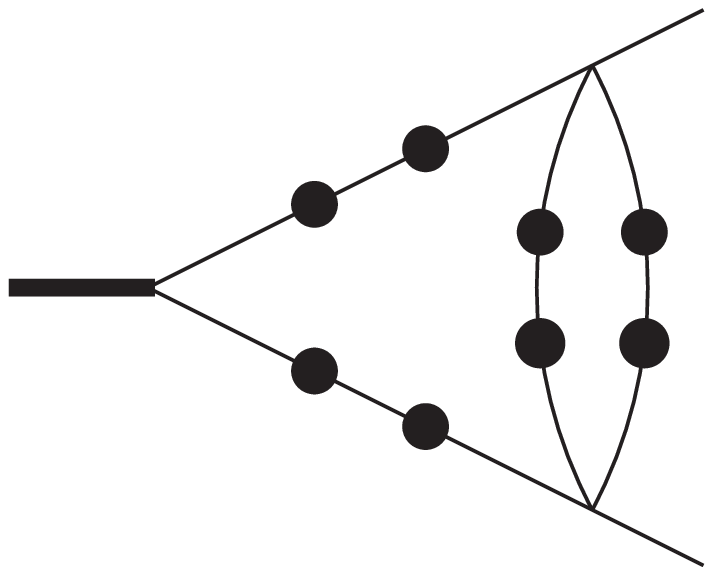}{10}\right\}
+ \frac{f_4}{\epsilon} \figgraph{.15}{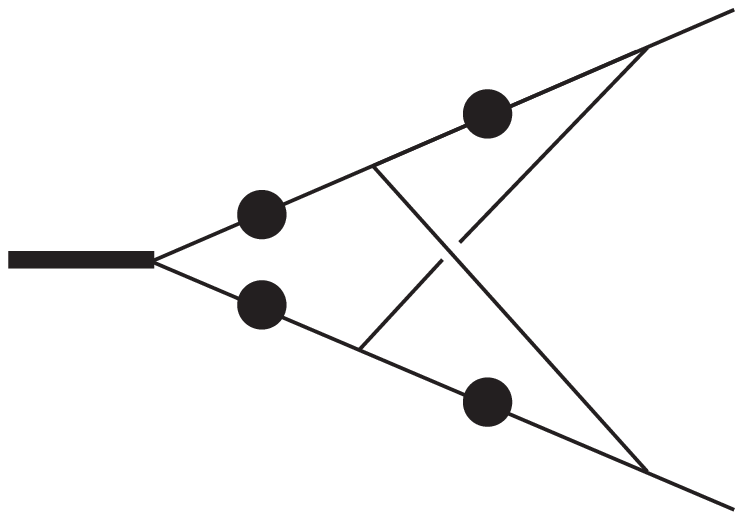}{8}\right)
\nonumber \\
&+ C_F C_A \left(\vphantom{\figgraph{.15}{triangle-bubble_finite}{10}}\frac{1}{\epsilon^4} \left[ f_5 \left\{-8 (1 - 3 \epsilon) (2 - 3 \epsilon)\figgraph{.15}{sunrise_finite}{8}\right\}
+f_6 \left\{\frac{32(1-2 \epsilon)(3-2 \epsilon)(1-3 \epsilon)(2-3 \epsilon)}{(1- \epsilon) (1+2 \epsilon)} \figgraph{.15}{triangle-bubble_finite}{10}\right\}\right] + \frac{f_7}{\epsilon} \figgraph{.15}{crossed_form_factor_finite}{8}\right)
\nonumber \\
& + C_F N_f \left(\frac{f_8}{\epsilon^3} \left\{\frac{32(1-2 \epsilon)(3-2 \epsilon)(1-3 \epsilon)(2-3 \epsilon)}{(1- \epsilon) (1+2 \epsilon)} \figgraph{.15}{triangle-bubble_finite}{10}\right\}\right). 
\end{align}
\end{widetext}
In Eq. (\ref{eq:badintegrand}), uniformly-finite products are enclosed with braces and the $f_i$ are both non-singular and non-vanishing in the $\epsilon \to 0$ limit. If desired, explicit expressions for the $f_i$ may be derived in a straightforward manner using the {\tt Reduze 2} implementation of Laporta's algorithm \cite{Laporta:2001dd} and the Tarasov dimension shift \cite{Tarasov:1996br,Lee:2010wea}, together with Eq. (10) of \cite{Gehrmann:2005pd}. Clearly, the expression for $\mathcal{F}_2^q(\epsilon)$ above sheds no light on its infrared structure, despite the fact that we have exhibited uniformly-finite integral candidates for all non-factorizable subtopologies.  

In fact, insisting upon the conventional master integral topologies is the problem here; a much more fruitful approach is to allow for integral topologies which are reducible with respect to integration by parts reduction. One can then perform a detailed scan of the available finite master integrals across all integral topologies using {\tt Reduze 2}. Two reducible topologies which stand out are
\begin{align}
\vcenter{\hbox{\includegraphics[scale=.2]{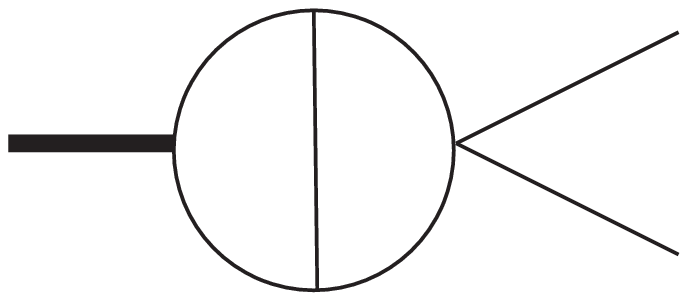}}} \qquad\qquad \mathrm{and}\qquad\qquad \vcenter{\hbox{\includegraphics[scale=.175]{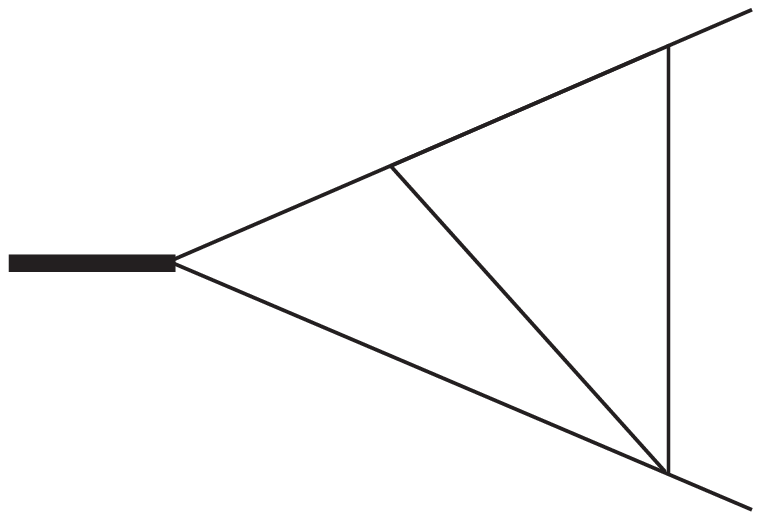}}}\, .\nonumber
\end{align}

What makes these topologies special is that they each admit just one integral candidate in the spacetime dimension where finite integrals first appear. For such distinguished finite integrals, we have seen in many examples that uniformly-finite integrals can be built up with dressing factors, even when there is nothing trivial about the topology in question. Although this trick turns out to be sufficient to find suitable replacements for the non-factorizable subtopologies in this case, one can expect in general that more analysis of the available integral topologies will be required, analogous to that which will be applied to the top-level, non-planar integral topology in what follows. 

The final ingredient required to fix $\epsilon$-dependent dressing factors is a framework for the evaluation of $d$-dimensional cut Feynman integrals \cite{Gehrmann-DeRidder:2003pne,Lee:2012te,Anastasiou:2013srw,Abreu:2014cla,Primo:2016ebd,Frellesvig:2017aai,Bosma:2017ens,Primo:2017ipr,Harley:2017qut,Lee:2017ftw}. Employing Eq. (2.13) of \cite{Harley:2017qut}, we find the uniformly-finite integrals
\begin{align}\label{eq:FUW_sunrise}
&-\frac{1}{6}(1-2\epsilon)\figgraph{.2}{eyeball_FUW}{4} = \zeta_3 + \frac{3}{5}\zeta_2^2 \epsilon 
+\left(-\zeta_2 \zeta_3 + 7 \zeta_5 \right)\epsilon^2 
\nonumber \\
&+ \left(\frac{99}{35}\zeta_2^3 - \frac{23}{3}\zeta_3^2\right) \epsilon^3
+\left(-\frac{17}{2}\zeta_2^2 \zeta_3 - 7 \zeta_2 \zeta_5 + 49 \zeta_7\right)\epsilon^4
\nonumber \\
& + \left(\frac{3777}{350}\zeta_2^4+\frac{23}{3} \zeta_2 \zeta_3^2-\frac{1306}{15}\zeta_3 \zeta_5\right)\epsilon^5 + \mathcal{O}\left(\epsilon^6\right)
\end{align}
and
\begin{align}\label{eq:FUW_triangle-bubble}
&-(1-2\epsilon)(1-3\epsilon)\figgraph{.175}{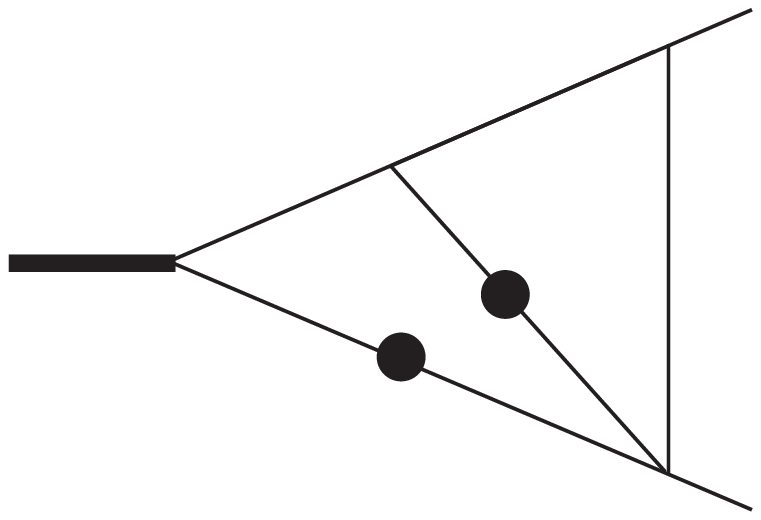}{6} = \zeta_2 + \zeta_3 \epsilon + \frac{4}{5}\zeta_2^2 \epsilon^2 
\nonumber \\
&+ \left(-\frac{29}{3}\zeta_2 \zeta_3 + 3 \zeta_5\right)\epsilon^3 + \left(-\frac{281}{70}\zeta_2^3 - \frac{29}{3}\zeta_3^2\right)\epsilon^4
\nonumber \\
&+\left(-\frac{379}{30}\zeta_2^2\zeta_3-\frac{257}{5}\zeta_2\zeta_5+9\zeta_7\right)\epsilon^5
\nonumber \\
&+\left(-\frac{4502}{175}\zeta_2^4+\frac{425}{9}\zeta_2\zeta_3^2-\frac{402}{5}\zeta_3\zeta_5\right)\epsilon^6 + \mathcal{O}\left(\epsilon^7\right).
\end{align}
To fix the dressing factors in the above, we found it natural to compute real-virtual two-propagator cuts. As desired, the improved basis integrals above do not contribute to the $\epsilon^{-4}$ or $\epsilon^{-3}$ pole terms of the amplitude.

One might think that such simple results arise because the integrals have an underlying representation in terms of Gamma functions. In contrast, the basic scalar  two-loop non-planar form factor integral was shown in \cite{Gehrmann:2005pd} to contain ${}_3F_2$ and ${}_4F_3$ functions in its all-orders-in-$\epsilon$ evaluation. It is therefore of some importance to show that we can readily construct a uniformly-finite integral candidate for the non-planar topology.

The non-planar topology's basic scalar integral in $4 - 2\epsilon$ is well-known to be uniform weight (see {\it e.g.} \cite{Smirnov:2004ym}). In general, however, the identification of uniform weight integrals in topologies which admit them is a non-trivial task and several programs which help one to accomplish this task have been published in the last few years \cite{Smirnov:2014hma,Maierhoefer:2017hyi,Prausa:2017ltv,Gituliar:2017vzm,Meyer:2017joq}. For form factors in $4 -2\epsilon$, a dedicated method has been developed \cite{Boels:2017skl,Boels:2017ftb}. If the result was not already known, the above-cited method would have allowed us to discover that the basic scalar integral in $4 - 2\epsilon$ is uniform weight with very little effort.

The non-planar topology first admits finite integrals in $6 - 2\epsilon$ and there are just two independent candidates,
\begin{align}
\figgraph{.175}{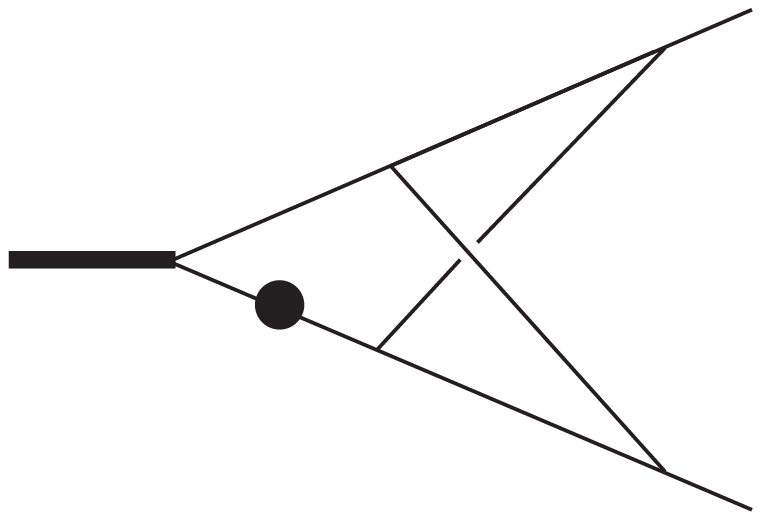}{6} \qquad\qquad \mathrm{and}\qquad\qquad \figgraph{.175}{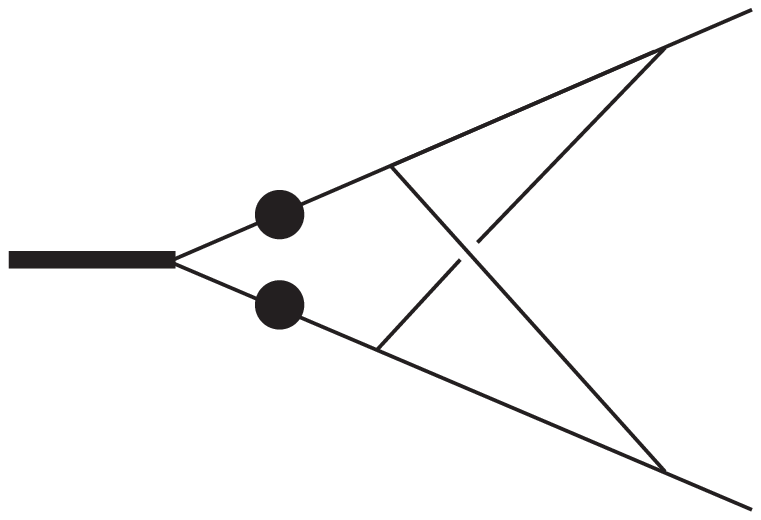}{6}\, .\nonumber
\end{align}
For this topology, a maximal cut analysis \cite{Bern:2004ky,Britto:2004nc,Cachazo:2008vp} of the finite integrals is not enough by itself, but it imposes strong constraints on the structure of the result. Eq. (2.13) of \cite{Harley:2017qut} implies
\begin{align}
\label{eq:max_cut_1}
&\figgraph{.175}{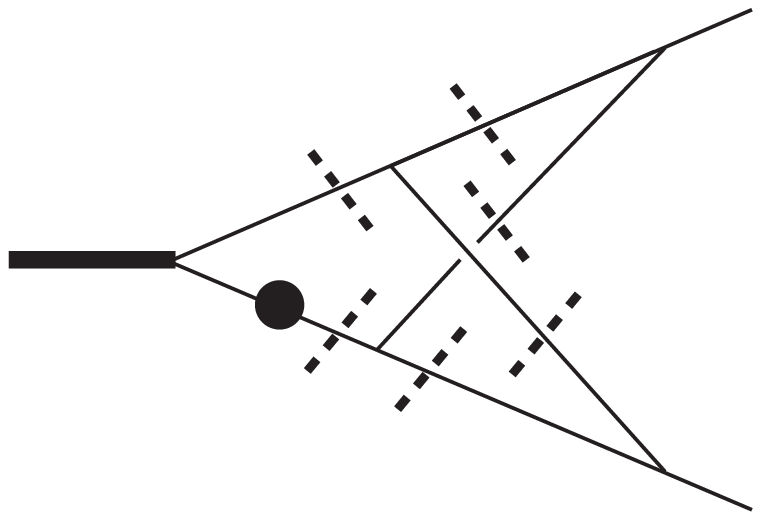}{6}=\frac{4 i \pi^3 e^{- i \pi \epsilon} e^{2 \epsilon \gamma_E}\Gamma(1-2\epsilon)}{(1-\epsilon)(1-4\epsilon)\Gamma(1-4\epsilon) \left(q^2\right)^{1+2\epsilon}}
\\
\label{eq:max_cut_2}
&\figgraph{.175}{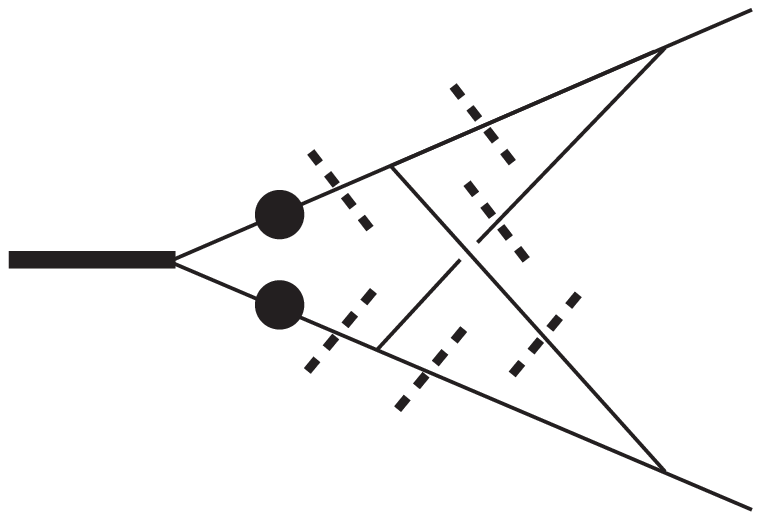}{6}=\frac{16 i \pi^3 e^{- i \pi \epsilon} e^{2 \epsilon \gamma_E}\epsilon \Gamma(1-2\epsilon)}{(1-\epsilon)(1-4\epsilon)\Gamma(1-4\epsilon) \left(q^2\right)^{2+2\epsilon}}
\end{align}
and we immediately see that the simplest linear combination of finite integrals which could turn out to simultaneously be uniform weight is of the form
\begin{align}
(1-a_1 \epsilon)(1-a_2 \epsilon)\figgraph{.175}{crossed_form_factor_FUW_1}{6} + a_3 (1-a_4 \epsilon)\figgraph{.175}{crossed_form_factor_FUW_2}{6} .\nonumber
\end{align}
Temporarily restoring the $q^2$ dependence, expanding the maximal cut of the above linear combination to $\mathcal{O}\left(\epsilon^3\right)$, and then demanding that all terms which do not have maximal weight cancel at each order in the $\epsilon$ expansion, we find that
\begin{align}
a_2 = 5-a_1-4 a_3 \quad\mathrm{and}\quad a_4= \frac{1-\frac{a_1}{4}(5 -a_1-4 a_3)}{a_3}\, .\nonumber
\end{align}

If we assume that the integration by parts reductions required to map onto a finite basis are available and that we have already explicitly evaluated all master integrals in subtopologies, we can fix the remaining constants by expressing the basic scalar integral in $4 - 2\epsilon$ as a linear combination of our putative uniformly-finite basis integrals. With the help of {\tt Reduze 2}, we find
\begin{align}
\label{eq:crossed_corner}
&\figgraph{.175}{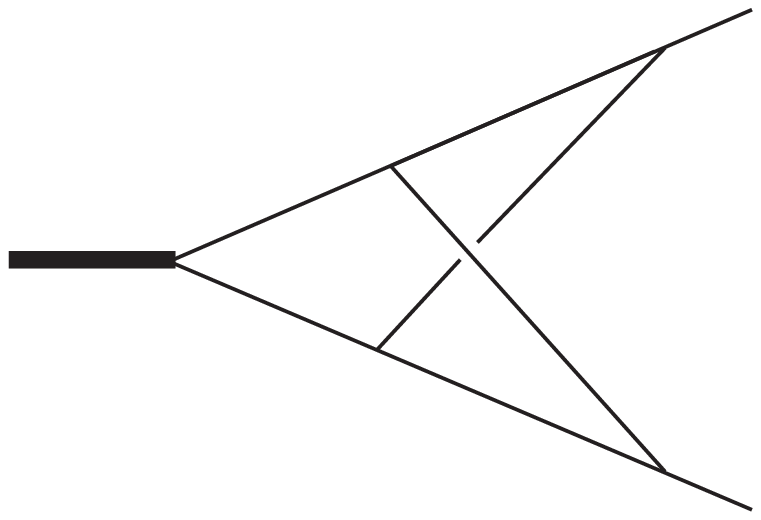}{4} = \frac{(1-2\epsilon)^2}{\epsilon^4}\figgraph{.175}{double-bubble_finite}{6}+\frac{g(a_1,a_3,\epsilon)}{\epsilon^2}\figgraph{.175}{triangle-bubble_FUW}{6}
\nonumber \\
& +\frac{1-2\epsilon}{\epsilon}\figgraph{.2}{eyeball_FUW}{4} + \frac{4}{\epsilon}\left\{(1-a_1 \epsilon)(1-a_2 \epsilon)\figgraph{.175}{crossed_form_factor_FUW_1}{6}
\right. \nonumber \\
& \left. + a_3 (1-a_4 \epsilon)\figgraph{.175}{crossed_form_factor_FUW_2}{6}\right\},
\end{align}
where 
\begin{align}
g(a_1,a_3,\epsilon) = 5 + (-21+8 a_3) \epsilon +(6 + 2 a_1 a_2) \epsilon^2 .\nonumber
\end{align}

The left-hand side of Eq. (\ref{eq:crossed_corner}) is known to be uniform weight and our remaining undetermined constants appear exclusively in the coefficient of integral (\ref{eq:FUW_triangle-bubble}). Therefore, we can fix $a_1$ and $a_3$ by expanding the second term on the right-hand side of Eq. (\ref{eq:crossed_corner}) to $\mathcal{O}\left(\epsilon^0\right)$ and then demanding that all terms proportional to $\zeta_2$ cancel at $\mathcal{O}\left(\epsilon^{-1}\right)$ and $\mathcal{O}\left(\epsilon^0\right)$. Going through these steps, we ultimately find the unique linear combination
\begin{align}
(1-4 \epsilon)(1-3 \epsilon)\figgraph{.175}{crossed_form_factor_FUW_1}{6} -\frac{1}{2} (1-4 \epsilon)\figgraph{.175}{crossed_form_factor_FUW_2}{6} .\nonumber
\end{align}

Finally, we can evaluate the underlying finite Feynman integrals using {\tt HyperInt} \cite{Panzer:2014caa,vonManteuffel:2015gxa} and normalize our newly-constructed integral as we did above, 
\begin{align}
&-\frac{1}{6}(1-4 \epsilon)\left(2 (1-3 \epsilon)\figgraph{.175}{crossed_form_factor_FUW_1}{6} - \figgraph{.175}{crossed_form_factor_FUW_2}{6}\right) = \zeta_3
\nonumber \\
& + \frac{2}{3}\zeta_2^2 \epsilon +\left(-\frac{5}{3}\zeta_2 \zeta_3+4\zeta_5\right)\epsilon^2+\left(-\frac{22}{35}\zeta_2^3-15\zeta_3^2\right)\epsilon^3
\nonumber \\
&+\left(-\frac{1681}{90}\zeta_2^2\zeta_3-4\zeta_2\zeta_5-\frac{127}{4}\zeta_7\right)\epsilon^4+\left(-\frac{1179}{35}\zeta_2^4
\right. \nonumber \\
&\left. +\frac{217}{9}\zeta_2\zeta_3^2-\frac{1966}{15}\zeta_3\zeta_5+22 \zeta_{5,3}\right)\epsilon^5 + \mathcal{O}\left(\epsilon^6\right)\,.
\end{align}
\section{discussion and outlook}
Adopting the uniformly-finite integral basis constructed in the previous section, our final result for the unrenormalized two-loop massless QCD quark form factor at virtuality $q^2 = -1$ has the schematic form
\begin{widetext}
\begin{align}
\label{eq:goodintegrand}
&\mathcal{F}_2^q(\epsilon) = C_F^2 \left(\frac{c_1}{\epsilon^4} \left\{(1-2\epsilon)^2 \figgraph{.15}{double-bubble_finite}{6}\right\} + \frac{1}{\epsilon} \left[c_2 \left\{-\frac{1}{6}(1-2\epsilon) \figgraph{.175}{eyeball_FUW}{4}\right\} 
+ c_3 \left\{-(1-2\epsilon)(1-3\epsilon) \figgraph{.15}{triangle-bubble_FUW}{6}\right\}
\right. \right. \nonumber \\& \left. \left.
+ c_4 \left\{-\frac{1}{6}(1-4\epsilon) \left(2 (1-3\epsilon) \figgraph{.15}{crossed_form_factor_FUW_1}{6} - \figgraph{.15}{crossed_form_factor_FUW_2}{6}\right)\right\}\right]\right) + C_F C_A \left(\frac{c_5}{\epsilon^3} \left\{(1-2\epsilon)^2 \figgraph{.15}{double-bubble_finite}{6}\right\}+\frac{c_6}{\epsilon^2} \times
\right. \nonumber \\& \left.
\times\left\{-(1-2\epsilon)(1-3\epsilon) \figgraph{.15}{triangle-bubble_FUW}{6}\right\}
+\frac{c_7}{\epsilon} \left\{-\frac{1}{6}(1-4\epsilon) \left(2(1-3\epsilon) \figgraph{.15}{crossed_form_factor_FUW_1}{6} - \figgraph{.15}{crossed_form_factor_FUW_2}{6}\right)\right\}
 + c_8 \left\{-\frac{1}{6}(1-2\epsilon)\figgraph{.175}{eyeball_FUW}{4}\right\}\right)
\nonumber \\&
+ C_F N_f \left(\frac{c_{9}}{\epsilon^3} \left\{(1-2\epsilon)^2 \figgraph{.15}{double-bubble_finite}{6}\right\}+\frac{c_{10}}{\epsilon} \left\{-(1-2\epsilon)(1-3\epsilon) \figgraph{.15}{triangle-bubble_FUW}{6}\right\} + c_{11} \left\{-\frac{1}{6}(1-2\epsilon) \figgraph{.175}{eyeball_FUW}{4}\right\}\right),
\end{align}
\end{widetext}
where the $c_i$ are both non-singular and non-vanishing in the $\epsilon \to 0$ limit.

Eq. (\ref{eq:goodintegrand}) makes manifest the connection between the one-loop form factor and the $\epsilon^{-4}$ and $\epsilon^{-3}$ poles of the two-loop form factor. Not only do these divergences now originate from a term proportional to the square of the one-loop master integral, we have furthermore eliminated a hidden zero that Eq. (\ref{eq:badintegrand}) had in the $\epsilon^{-4}$ term of its $C_A C_F$ color structure. Even better, the impact of the non-Abelian exponentiation theorem \cite{Gatheral:1983cz,Frenkel:1984pz} on the $C_F^2$ color structure is manifest now as well: we see from Eq. (\ref{eq:goodintegrand}) that the square of the dressed one-loop bubble generates even the $\epsilon^{-2}$ divergence of $C_F^2$.

In this paper, we introduced a new Feynman integral basis which unifies the well-known finite and uniform weight integral basis paradigms. We have seen that, by making judicious choices for the basis elements, a finite and uniform weight basis can expose structure in the integrands of phenomenologically-relevant scattering amplitudes which has until now remained completely hidden from view. Furthermore, our explicit construction was shown to require little effort beyond that which is needed for the setup of an ordinary uniform weight basis.

Of course, many interesting questions remain. As mentioned in the introduction, it is natural to wonder whether a similar construction could work in the presence of massive quarks. A first test would be to rerun the analysis of this paper for the two-loop heavy quark form factor in QCD \cite{Bernreuther:2004ih,Bernreuther:2004th,Bernreuther:2005rw,Bernreuther:2005gw,Gluza:2009yy,Ablinger:2017hst}. It is also important to continue studying scattering amplitudes in massless QCD, both with more legs and more loops. In this direction, we have already obtained a number of non-trivial results, but it seems clear that an analysis of the full three-loop quark form factor of massless QCD \cite{Moch:2005tm,Gehrmann:2006wg,Heinrich:2007at,Heinrich:2009be,Baikov:2009bg,Baikov:2009bg,Lee:2010cga,Gehrmann:2010ue} would be the next logical step to shed additional light on our proposal for the manifest exponentiation of the higher poles in $\epsilon$.

Naively, it would seem that we already have a problem, due to the fact that our one-loop form factor master lives in $6-2\epsilon$, but we chose the $4 - 2\epsilon$ kite integral as one of our two-loop form factor basis integrals. Fortunately, it turns out \cite{RobSCETtalk} that one can lift the $4 - 2\epsilon$ kite integral to a dressed sum of two $6 - 2\epsilon$ planar ladder integrals:
\begin{equation}
\figgraph{.195}{eyeball_FUW}{4} = 2 (1-2\epsilon) \left(\figgraph{.195}{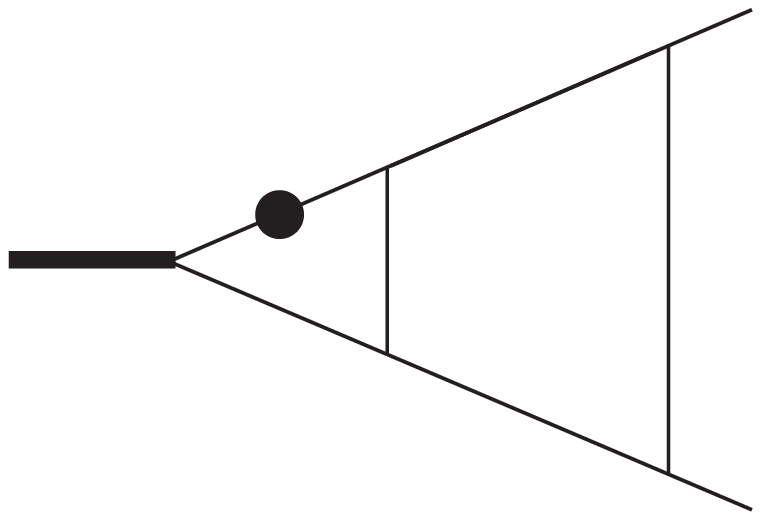}{6} 
+ \figgraph{.195}{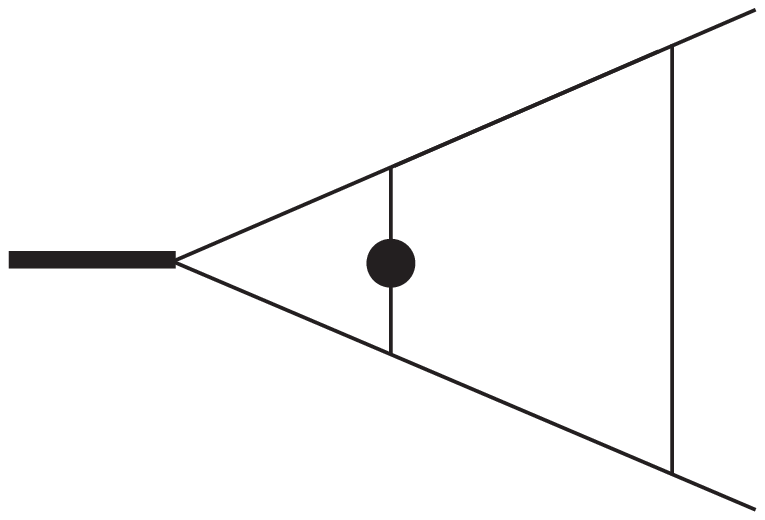}{6}\right).
\end{equation}
Clearly, it would be useful to be able to predict in advance which integral topologies mesh well with our framework.

Hopefully, our construction will also provide another direct link between the study of scattering amplitudes in $\mathcal{N} = 4$ super Yang-Mills theory \cite{Brink:1976bc} and more realistic models like QCD. Research into $\mathcal{N} = 4$ super Yang-Mills theory scattering amplitudes has begun to branch out to the non-planar sector \cite{Arkani-Hamed:2014bca,Bern:2015ple,Bourjaily:2016mnp,Bern:2018oao} and it seems possible that our program  could benefit from various recent discoveries. One idea would be to investigate whether there is deep mathematics behind the question of why certain topologies admit finite and uniform weight integrals with high leading weights and others do not.
\vspace{.7 cm}

\noindent {\em Note Added:}
After this work was accepted for publication, we realized that an interesting example of the ideas we presented was put forth already in \cite{Caron-Huot:2014lda}, albeit in the context of light-by-light scattering in planar $\mathcal{N} = 4$ super Yang-Mills theory \cite{Kubo:1984hm}; by regulating the infrared with a vacuum expectation value for one of the scalar fields, all Feynman integrals contributing to scattering amplitudes in the model are automatically rendered finite in four spacetime dimensions (see {\it e.g.} \cite{Schabinger:2008ah,Alday:2009zm,Boels:2010mj}).

\noindent {\em Acknowledgments:}
RMS gratefully acknowledges inspiring discussions with Claude Duhr and Bernhard Mistlberger at Amplitudes 2017 and useful discussions with the Trinity College Dublin amplitudes group at an early stage of this work. We are particularly indebted to Ruth Britto and Andreas von Manteuffel for their advice and encouragement while this work was being carried out. This research was supported in part by the European Research Council through grant 647356 (CutLoops). Our figures were generated using {\tt Jaxodraw} \cite{Binosi:2003yf}, based on {\tt AxoDraw} \cite{Vermaseren:1994je}.

\bibliography{uniformlyfinite}

\begin{thebibliography}{126}%
\makeatletter
\providecommand \@ifxundefined [1]{%
 \@ifx{#1\undefined}
}%
\providecommand \@ifnum [1]{%
 \ifnum #1\expandafter \@firstoftwo
 \else \expandafter \@secondoftwo
 \fi
}%
\providecommand \@ifx [1]{%
 \ifx #1\expandafter \@firstoftwo
 \else \expandafter \@secondoftwo
 \fi
}%
\providecommand \natexlab [1]{#1}%
\providecommand \enquote  [1]{``#1''}%
\providecommand \bibnamefont  [1]{#1}%
\providecommand \bibfnamefont [1]{#1}%
\providecommand \citenamefont [1]{#1}%
\providecommand \href@noop [0]{\@secondoftwo}%
\providecommand \href [0]{\begingroup \@sanitize@url \@href}%
\providecommand \@href[1]{\@@startlink{#1}\@@href}%
\providecommand \@@href[1]{\endgroup#1\@@endlink}%
\providecommand \@sanitize@url [0]{\catcode `\\12\catcode `\$12\catcode
  `\&12\catcode `\#12\catcode `\^12\catcode `\_12\catcode `\%12\relax}%
\providecommand \@@startlink[1]{}%
\providecommand \@@endlink[0]{}%
\providecommand \url  [0]{\begingroup\@sanitize@url \@url }%
\providecommand \@url [1]{\endgroup\@href {#1}{\urlprefix }}%
\providecommand \urlprefix  [0]{URL }%
\providecommand \Eprint [0]{\href }%
\providecommand \doibase [0]{http://dx.doi.org/}%
\providecommand \selectlanguage [0]{\@gobble}%
\providecommand \bibinfo  [0]{\@secondoftwo}%
\providecommand \bibfield  [0]{\@secondoftwo}%
\providecommand \translation [1]{[#1]}%
\providecommand \BibitemOpen [0]{}%
\providecommand \bibitemStop [0]{}%
\providecommand \bibitemNoStop [0]{.\EOS\space}%
\providecommand \EOS [0]{\spacefactor3000\relax}%
\providecommand \BibitemShut  [1]{\csname bibitem#1\endcsname}%
\let\auto@bib@innerbib\@empty
\bibitem [{\citenamefont {Brandt}\ \emph {et~al.}(1982)\citenamefont {Brandt},
  \citenamefont {Gocksch}, \citenamefont {Sato},\ and\ \citenamefont
  {Neri}}]{Brandt:1982gz}%
  \BibitemOpen
  \bibfield  {author} {\bibinfo {author} {\bibfnamefont {R.~A.}\ \bibnamefont
  {Brandt}}, \bibinfo {author} {\bibfnamefont {A.}~\bibnamefont {Gocksch}},
  \bibinfo {author} {\bibfnamefont {M.~A.}\ \bibnamefont {Sato}}, \ and\
  \bibinfo {author} {\bibfnamefont {F.}~\bibnamefont {Neri}},\ }\href {\doibase
  10.1103/PhysRevD.26.3611} {\bibfield  {journal} {\bibinfo  {journal} {Phys.
  Rev.}\ }\textbf {\bibinfo {volume} {D26}},\ \bibinfo {pages} {3611} (\bibinfo
  {year} {1982})}\BibitemShut {NoStop}%
\bibitem [{\citenamefont {Gatheral}(1983)}]{Gatheral:1983cz}%
  \BibitemOpen
  \bibfield  {author} {\bibinfo {author} {\bibfnamefont {J.~G.~M.}\
  \bibnamefont {Gatheral}},\ }\href {\doibase 10.1016/0370-2693(83)90112-0}
  {\bibfield  {journal} {\bibinfo  {journal} {Phys. Lett.}\ }\textbf {\bibinfo
  {volume} {133B}},\ \bibinfo {pages} {90} (\bibinfo {year}
  {1983})}\BibitemShut {NoStop}%
\bibitem [{\citenamefont {Frenkel}\ and\ \citenamefont
  {Taylor}(1984)}]{Frenkel:1984pz}%
  \BibitemOpen
  \bibfield  {author} {\bibinfo {author} {\bibfnamefont {J.}~\bibnamefont
  {Frenkel}}\ and\ \bibinfo {author} {\bibfnamefont {J.~C.}\ \bibnamefont
  {Taylor}},\ }\href {\doibase 10.1016/0550-3213(84)90294-3} {\bibfield
  {journal} {\bibinfo  {journal} {Nucl. Phys.}\ }\textbf {\bibinfo {volume}
  {B246}},\ \bibinfo {pages} {231} (\bibinfo {year} {1984})}\BibitemShut
  {NoStop}%
\bibitem [{\citenamefont {Korchemsky}\ and\ \citenamefont
  {Radyushkin}(1986)}]{Korchemsky:1985xj}%
  \BibitemOpen
  \bibfield  {author} {\bibinfo {author} {\bibfnamefont {G.~P.}\ \bibnamefont
  {Korchemsky}}\ and\ \bibinfo {author} {\bibfnamefont {A.~V.}\ \bibnamefont
  {Radyushkin}},\ }\href {\doibase 10.1016/0370-2693(86)91439-5} {\bibfield
  {journal} {\bibinfo  {journal} {Phys. Lett.}\ }\textbf {\bibinfo {volume}
  {B171}},\ \bibinfo {pages} {459} (\bibinfo {year} {1986})}\BibitemShut
  {NoStop}%
\bibitem [{\citenamefont {Sterman}(1987)}]{Sterman:1986aj}%
  \BibitemOpen
  \bibfield  {author} {\bibinfo {author} {\bibfnamefont {G.~F.}\ \bibnamefont
  {Sterman}},\ }\href {\doibase 10.1016/0550-3213(87)90258-6} {\bibfield
  {journal} {\bibinfo  {journal} {Nucl. Phys.}\ }\textbf {\bibinfo {volume}
  {B281}},\ \bibinfo {pages} {310} (\bibinfo {year} {1987})}\BibitemShut
  {NoStop}%
\bibitem [{\citenamefont {Korchemsky}\ and\ \citenamefont
  {Radyushkin}(1987)}]{Korchemsky:1987wg}%
  \BibitemOpen
  \bibfield  {author} {\bibinfo {author} {\bibfnamefont {G.~P.}\ \bibnamefont
  {Korchemsky}}\ and\ \bibinfo {author} {\bibfnamefont {A.~V.}\ \bibnamefont
  {Radyushkin}},\ }\href {\doibase 10.1016/0550-3213(87)90277-X} {\bibfield
  {journal} {\bibinfo  {journal} {Nucl. Phys.}\ }\textbf {\bibinfo {volume}
  {B283}},\ \bibinfo {pages} {342} (\bibinfo {year} {1987})}\BibitemShut
  {NoStop}%
\bibitem [{\citenamefont {Collins}\ \emph {et~al.}(1989)\citenamefont
  {Collins}, \citenamefont {Soper},\ and\ \citenamefont
  {Sterman}}]{Collins:1989gx}%
  \BibitemOpen
  \bibfield  {author} {\bibinfo {author} {\bibfnamefont {J.~C.}\ \bibnamefont
  {Collins}}, \bibinfo {author} {\bibfnamefont {D.~E.}\ \bibnamefont {Soper}},
  \ and\ \bibinfo {author} {\bibfnamefont {G.~F.}\ \bibnamefont {Sterman}},\
  }\href {\doibase 10.1142/9789814503266_0001} {\bibfield  {journal} {\bibinfo
  {journal} {Adv. Ser. Direct. High Energy Phys.}\ }\textbf {\bibinfo {volume}
  {5}},\ \bibinfo {pages} {1} (\bibinfo {year} {1989})},\ \Eprint
  {http://arxiv.org/abs/hep-ph/0409313} {arXiv:hep-ph/0409313} \BibitemShut
  {NoStop}%
\bibitem [{\citenamefont {Korchemsky}(1989{\natexlab{a}})}]{Korchemsky:1988hd}%
  \BibitemOpen
  \bibfield  {author} {\bibinfo {author} {\bibfnamefont {G.~P.}\ \bibnamefont
  {Korchemsky}},\ }\href {\doibase 10.1016/0370-2693(89)90799-5} {\bibfield
  {journal} {\bibinfo  {journal} {Phys. Lett.}\ }\textbf {\bibinfo {volume}
  {B220}},\ \bibinfo {pages} {629} (\bibinfo {year}
  {1989}{\natexlab{a}})}\BibitemShut {NoStop}%
\bibitem [{\citenamefont {Korchemsky}(1989{\natexlab{b}})}]{Korchemsky:1988si}%
  \BibitemOpen
  \bibfield  {author} {\bibinfo {author} {\bibfnamefont {G.~P.}\ \bibnamefont
  {Korchemsky}},\ }\href {\doibase 10.1142/S0217732389001453} {\bibfield
  {journal} {\bibinfo  {journal} {Mod. Phys. Lett.}\ }\textbf {\bibinfo
  {volume} {A4}},\ \bibinfo {pages} {1257} (\bibinfo {year}
  {1989}{\natexlab{b}})}\BibitemShut {NoStop}%
\bibitem [{\citenamefont {Magnea}\ and\ \citenamefont
  {Sterman}(1990)}]{Magnea:1990zb}%
  \BibitemOpen
  \bibfield  {author} {\bibinfo {author} {\bibfnamefont {L.}~\bibnamefont
  {Magnea}}\ and\ \bibinfo {author} {\bibfnamefont {G.~F.}\ \bibnamefont
  {Sterman}},\ }\href {\doibase 10.1103/PhysRevD.42.4222} {\bibfield  {journal}
  {\bibinfo  {journal} {Phys. Rev.}\ }\textbf {\bibinfo {volume} {D42}},\
  \bibinfo {pages} {4222} (\bibinfo {year} {1990})}\BibitemShut {NoStop}%
\bibitem [{\citenamefont {Korchemskaya}\ and\ \citenamefont
  {Korchemsky}(1995)}]{Korchemskaya:1994qp}%
  \BibitemOpen
  \bibfield  {author} {\bibinfo {author} {\bibfnamefont {I.~A.}\ \bibnamefont
  {Korchemskaya}}\ and\ \bibinfo {author} {\bibfnamefont {G.~P.}\ \bibnamefont
  {Korchemsky}},\ }\href {\doibase 10.1016/0550-3213(94)00553-Q} {\bibfield
  {journal} {\bibinfo  {journal} {Nucl. Phys.}\ }\textbf {\bibinfo {volume}
  {B437}},\ \bibinfo {pages} {127} (\bibinfo {year} {1995})},\ \Eprint
  {http://arxiv.org/abs/hep-ph/9409446} {arXiv:hep-ph/9409446} \BibitemShut
  {NoStop}%
\bibitem [{\citenamefont {Catani}\ and\ \citenamefont
  {Seymour}(1996)}]{Catani:1996jh}%
  \BibitemOpen
  \bibfield  {author} {\bibinfo {author} {\bibfnamefont {S.}~\bibnamefont
  {Catani}}\ and\ \bibinfo {author} {\bibfnamefont {M.~H.}\ \bibnamefont
  {Seymour}},\ }\href {\doibase 10.1016/0370-2693(96)00425-X} {\bibfield
  {journal} {\bibinfo  {journal} {Phys. Lett.}\ }\textbf {\bibinfo {volume}
  {B378}},\ \bibinfo {pages} {287} (\bibinfo {year} {1996})},\ \Eprint
  {http://arxiv.org/abs/hep-ph/9602277} {arXiv:hep-ph/9602277} \BibitemShut
  {NoStop}%
\bibitem [{\citenamefont {Contopanagos}\ \emph {et~al.}(1997)\citenamefont
  {Contopanagos}, \citenamefont {Laenen},\ and\ \citenamefont
  {Sterman}}]{Contopanagos:1996nh}%
  \BibitemOpen
  \bibfield  {author} {\bibinfo {author} {\bibfnamefont {H.}~\bibnamefont
  {Contopanagos}}, \bibinfo {author} {\bibfnamefont {E.}~\bibnamefont
  {Laenen}}, \ and\ \bibinfo {author} {\bibfnamefont {G.~F.}\ \bibnamefont
  {Sterman}},\ }\href {\doibase 10.1016/S0550-3213(96)00567-6} {\bibfield
  {journal} {\bibinfo  {journal} {Nucl. Phys.}\ }\textbf {\bibinfo {volume}
  {B484}},\ \bibinfo {pages} {303} (\bibinfo {year} {1997})},\ \Eprint
  {http://arxiv.org/abs/hep-ph/9604313} {arXiv:hep-ph/9604313} \BibitemShut
  {NoStop}%
\bibitem [{\citenamefont {Kidonakis}\ and\ \citenamefont
  {Sterman}(1997)}]{Kidonakis:1997gm}%
  \BibitemOpen
  \bibfield  {author} {\bibinfo {author} {\bibfnamefont {N.}~\bibnamefont
  {Kidonakis}}\ and\ \bibinfo {author} {\bibfnamefont {G.~F.}\ \bibnamefont
  {Sterman}},\ }\href {\doibase 10.1016/S0550-3213(97)00506-3} {\bibfield
  {journal} {\bibinfo  {journal} {Nucl. Phys.}\ }\textbf {\bibinfo {volume}
  {B505}},\ \bibinfo {pages} {321} (\bibinfo {year} {1997})},\ \Eprint
  {http://arxiv.org/abs/hep-ph/9705234} {arXiv:hep-ph/9705234} \BibitemShut
  {NoStop}%
\bibitem [{\citenamefont {Kidonakis}\ \emph
  {et~al.}(1998{\natexlab{a}})\citenamefont {Kidonakis}, \citenamefont
  {Oderda},\ and\ \citenamefont {Sterman}}]{Kidonakis:1998bk}%
  \BibitemOpen
  \bibfield  {author} {\bibinfo {author} {\bibfnamefont {N.}~\bibnamefont
  {Kidonakis}}, \bibinfo {author} {\bibfnamefont {G.}~\bibnamefont {Oderda}}, \
  and\ \bibinfo {author} {\bibfnamefont {G.~F.}\ \bibnamefont {Sterman}},\
  }\href {\doibase 10.1016/S0550-3213(98)00243-0} {\bibfield  {journal}
  {\bibinfo  {journal} {Nucl. Phys.}\ }\textbf {\bibinfo {volume} {B525}},\
  \bibinfo {pages} {299} (\bibinfo {year} {1998}{\natexlab{a}})},\ \Eprint
  {http://arxiv.org/abs/hep-ph/9801268} {arXiv:hep-ph/9801268} \BibitemShut
  {NoStop}%
\bibitem [{\citenamefont {Catani}(1998)}]{Catani:1998bh}%
  \BibitemOpen
  \bibfield  {author} {\bibinfo {author} {\bibfnamefont {S.}~\bibnamefont
  {Catani}},\ }\href {\doibase 10.1016/S0370-2693(98)00332-3} {\bibfield
  {journal} {\bibinfo  {journal} {Phys. Lett.}\ }\textbf {\bibinfo {volume}
  {B427}},\ \bibinfo {pages} {161} (\bibinfo {year} {1998})},\ \Eprint
  {http://arxiv.org/abs/hep-ph/9802439} {arXiv:hep-ph/9802439} \BibitemShut
  {NoStop}%
\bibitem [{\citenamefont {Kidonakis}\ \emph
  {et~al.}(1998{\natexlab{b}})\citenamefont {Kidonakis}, \citenamefont
  {Oderda},\ and\ \citenamefont {Sterman}}]{Kidonakis:1998nf}%
  \BibitemOpen
  \bibfield  {author} {\bibinfo {author} {\bibfnamefont {N.}~\bibnamefont
  {Kidonakis}}, \bibinfo {author} {\bibfnamefont {G.}~\bibnamefont {Oderda}}, \
  and\ \bibinfo {author} {\bibfnamefont {G.~F.}\ \bibnamefont {Sterman}},\
  }\href {\doibase 10.1016/S0550-3213(98)00441-6} {\bibfield  {journal}
  {\bibinfo  {journal} {Nucl. Phys.}\ }\textbf {\bibinfo {volume} {B531}},\
  \bibinfo {pages} {365} (\bibinfo {year} {1998}{\natexlab{b}})},\ \Eprint
  {http://arxiv.org/abs/hep-ph/9803241} {arXiv:hep-ph/9803241} \BibitemShut
  {NoStop}%
\bibitem [{\citenamefont {Sterman}\ and\ \citenamefont
  {Tejeda-Yeomans}(2003)}]{Sterman:2002qn}%
  \BibitemOpen
  \bibfield  {author} {\bibinfo {author} {\bibfnamefont {G.~F.}\ \bibnamefont
  {Sterman}}\ and\ \bibinfo {author} {\bibfnamefont {M.~E.}\ \bibnamefont
  {Tejeda-Yeomans}},\ }\href {\doibase 10.1016/S0370-2693(02)03100-3}
  {\bibfield  {journal} {\bibinfo  {journal} {Phys. Lett.}\ }\textbf {\bibinfo
  {volume} {B552}},\ \bibinfo {pages} {48} (\bibinfo {year} {2003})},\ \Eprint
  {http://arxiv.org/abs/hep-ph/0210130} {arXiv:hep-ph/0210130} \BibitemShut
  {NoStop}%
\bibitem [{\citenamefont {Dokshitzer}\ and\ \citenamefont
  {Marchesini}(2006)}]{Dokshitzer:2005ig}%
  \BibitemOpen
  \bibfield  {author} {\bibinfo {author} {\bibfnamefont {{\relax Yu}.~L.}\
  \bibnamefont {Dokshitzer}}\ and\ \bibinfo {author} {\bibfnamefont
  {G.}~\bibnamefont {Marchesini}},\ }\href {\doibase
  10.1088/1126-6708/2006/01/007} {\bibfield  {journal} {\bibinfo  {journal}
  {JHEP}\ }\textbf {\bibinfo {volume} {01}},\ \bibinfo {pages} {007} (\bibinfo
  {year} {2006})},\ \Eprint {http://arxiv.org/abs/hep-ph/0509078}
  {arXiv:hep-ph/0509078} \BibitemShut {NoStop}%
\bibitem [{\citenamefont {Aybat}\ \emph
  {et~al.}(2006{\natexlab{a}})\citenamefont {Aybat}, \citenamefont {Dixon},\
  and\ \citenamefont {Sterman}}]{Aybat:2006wq}%
  \BibitemOpen
  \bibfield  {author} {\bibinfo {author} {\bibfnamefont {S.~M.}\ \bibnamefont
  {Aybat}}, \bibinfo {author} {\bibfnamefont {L.~J.}\ \bibnamefont {Dixon}}, \
  and\ \bibinfo {author} {\bibfnamefont {G.~F.}\ \bibnamefont {Sterman}},\
  }\href {\doibase 10.1103/PhysRevLett.97.072001} {\bibfield  {journal}
  {\bibinfo  {journal} {Phys. Rev. Lett.}\ }\textbf {\bibinfo {volume} {97}},\
  \bibinfo {pages} {072001} (\bibinfo {year} {2006}{\natexlab{a}})},\ \Eprint
  {http://arxiv.org/abs/hep-ph/0606254} {arXiv:hep-ph/0606254} \BibitemShut
  {NoStop}%
\bibitem [{\citenamefont {Aybat}\ \emph
  {et~al.}(2006{\natexlab{b}})\citenamefont {Aybat}, \citenamefont {Dixon},\
  and\ \citenamefont {Sterman}}]{Aybat:2006mz}%
  \BibitemOpen
  \bibfield  {author} {\bibinfo {author} {\bibfnamefont {S.~M.}\ \bibnamefont
  {Aybat}}, \bibinfo {author} {\bibfnamefont {L.~J.}\ \bibnamefont {Dixon}}, \
  and\ \bibinfo {author} {\bibfnamefont {G.~F.}\ \bibnamefont {Sterman}},\
  }\href {\doibase 10.1103/PhysRevD.74.074004} {\bibfield  {journal} {\bibinfo
  {journal} {Phys. Rev.}\ }\textbf {\bibinfo {volume} {D74}},\ \bibinfo {pages}
  {074004} (\bibinfo {year} {2006}{\natexlab{b}})},\ \Eprint
  {http://arxiv.org/abs/hep-ph/0607309} {arXiv:hep-ph/0607309} \BibitemShut
  {NoStop}%
\bibitem [{\citenamefont {Mitov}\ and\ \citenamefont
  {Moch}(2007)}]{Mitov:2006xs}%
  \BibitemOpen
  \bibfield  {author} {\bibinfo {author} {\bibfnamefont {A.}~\bibnamefont
  {Mitov}}\ and\ \bibinfo {author} {\bibfnamefont {S.}~\bibnamefont {Moch}},\
  }\href {\doibase 10.1088/1126-6708/2007/05/001} {\bibfield  {journal}
  {\bibinfo  {journal} {JHEP}\ }\textbf {\bibinfo {volume} {05}},\ \bibinfo
  {pages} {001} (\bibinfo {year} {2007})},\ \Eprint
  {http://arxiv.org/abs/hep-ph/0612149} {arXiv:hep-ph/0612149} \BibitemShut
  {NoStop}%
\bibitem [{\citenamefont {Dixon}\ \emph {et~al.}(2008)\citenamefont {Dixon},
  \citenamefont {Magnea},\ and\ \citenamefont {Sterman}}]{Dixon:2008gr}%
  \BibitemOpen
  \bibfield  {author} {\bibinfo {author} {\bibfnamefont {L.~J.}\ \bibnamefont
  {Dixon}}, \bibinfo {author} {\bibfnamefont {L.}~\bibnamefont {Magnea}}, \
  and\ \bibinfo {author} {\bibfnamefont {G.~F.}\ \bibnamefont {Sterman}},\
  }\href {\doibase 10.1088/1126-6708/2008/08/022} {\bibfield  {journal}
  {\bibinfo  {journal} {JHEP}\ }\textbf {\bibinfo {volume} {08}},\ \bibinfo
  {pages} {022} (\bibinfo {year} {2008})},\ \Eprint
  {http://arxiv.org/abs/0805.3515} {arXiv:0805.3515} \BibitemShut {NoStop}%
\bibitem [{\citenamefont {Becher}\ and\ \citenamefont
  {Neubert}(2009{\natexlab{a}})}]{Becher:2009cu}%
  \BibitemOpen
  \bibfield  {author} {\bibinfo {author} {\bibfnamefont {T.}~\bibnamefont
  {Becher}}\ and\ \bibinfo {author} {\bibfnamefont {M.}~\bibnamefont
  {Neubert}},\ }\href {\doibase 10.1103/PhysRevLett.102.162001,
  10.1103/PhysRevLett.111.199905} {\bibfield  {journal} {\bibinfo  {journal}
  {Phys. Rev. Lett.}\ }\textbf {\bibinfo {volume} {102}},\ \bibinfo {pages}
  {162001} (\bibinfo {year} {2009}{\natexlab{a}})},\ \Eprint
  {http://arxiv.org/abs/0901.0722} {arXiv:0901.0722} \BibitemShut {NoStop}%
\bibitem [{\citenamefont {Gardi}\ and\ \citenamefont
  {Magnea}(2009)}]{Gardi:2009qi}%
  \BibitemOpen
  \bibfield  {author} {\bibinfo {author} {\bibfnamefont {E.}~\bibnamefont
  {Gardi}}\ and\ \bibinfo {author} {\bibfnamefont {L.}~\bibnamefont {Magnea}},\
  }\href {\doibase 10.1088/1126-6708/2009/03/079} {\bibfield  {journal}
  {\bibinfo  {journal} {JHEP}\ }\textbf {\bibinfo {volume} {03}},\ \bibinfo
  {pages} {079} (\bibinfo {year} {2009})},\ \Eprint
  {http://arxiv.org/abs/0901.1091} {arXiv:0901.1091} \BibitemShut {NoStop}%
\bibitem [{\citenamefont {Becher}\ and\ \citenamefont
  {Neubert}(2009{\natexlab{b}})}]{Becher:2009qa}%
  \BibitemOpen
  \bibfield  {author} {\bibinfo {author} {\bibfnamefont {T.}~\bibnamefont
  {Becher}}\ and\ \bibinfo {author} {\bibfnamefont {M.}~\bibnamefont
  {Neubert}},\ }\href {\doibase 10.1088/1126-6708/2009/06/081,
  10.1007/JHEP11(2013)024} {\bibfield  {journal} {\bibinfo  {journal} {JHEP}\
  }\textbf {\bibinfo {volume} {06}},\ \bibinfo {pages} {081} (\bibinfo {year}
  {2009}{\natexlab{b}})},\ \Eprint {http://arxiv.org/abs/0903.1126}
  {arXiv:0903.1126} \BibitemShut {NoStop}%
\bibitem [{\citenamefont {Becher}\ and\ \citenamefont
  {Neubert}(2009{\natexlab{c}})}]{Becher:2009kw}%
  \BibitemOpen
  \bibfield  {author} {\bibinfo {author} {\bibfnamefont {T.}~\bibnamefont
  {Becher}}\ and\ \bibinfo {author} {\bibfnamefont {M.}~\bibnamefont
  {Neubert}},\ }\href {\doibase 10.1103/PhysRevD.79.125004,
  10.1103/PhysRevD.80.109901} {\bibfield  {journal} {\bibinfo  {journal} {Phys.
  Rev.}\ }\textbf {\bibinfo {volume} {D79}},\ \bibinfo {pages} {125004}
  (\bibinfo {year} {2009}{\natexlab{c}})},\ \Eprint
  {http://arxiv.org/abs/0904.1021} {arXiv:0904.1021} \BibitemShut {NoStop}%
\bibitem [{\citenamefont {Dixon}\ \emph {et~al.}(2010)\citenamefont {Dixon},
  \citenamefont {Gardi},\ and\ \citenamefont {Magnea}}]{Dixon:2009ur}%
  \BibitemOpen
  \bibfield  {author} {\bibinfo {author} {\bibfnamefont {L.~J.}\ \bibnamefont
  {Dixon}}, \bibinfo {author} {\bibfnamefont {E.}~\bibnamefont {Gardi}}, \ and\
  \bibinfo {author} {\bibfnamefont {L.}~\bibnamefont {Magnea}},\ }\href
  {\doibase 10.1007/JHEP02(2010)081} {\bibfield  {journal} {\bibinfo  {journal}
  {JHEP}\ }\textbf {\bibinfo {volume} {02}},\ \bibinfo {pages} {081} (\bibinfo
  {year} {2010})},\ \Eprint {http://arxiv.org/abs/0910.3653} {arXiv:0910.3653}
  \BibitemShut {NoStop}%
\bibitem [{\citenamefont {Gardi}\ \emph {et~al.}(2010)\citenamefont {Gardi},
  \citenamefont {Laenen}, \citenamefont {Stavenga},\ and\ \citenamefont
  {White}}]{Gardi:2010rn}%
  \BibitemOpen
  \bibfield  {author} {\bibinfo {author} {\bibfnamefont {E.}~\bibnamefont
  {Gardi}}, \bibinfo {author} {\bibfnamefont {E.}~\bibnamefont {Laenen}},
  \bibinfo {author} {\bibfnamefont {G.}~\bibnamefont {Stavenga}}, \ and\
  \bibinfo {author} {\bibfnamefont {C.~D.}\ \bibnamefont {White}},\ }\href
  {\doibase 10.1007/JHEP11(2010)155} {\bibfield  {journal} {\bibinfo  {journal}
  {JHEP}\ }\textbf {\bibinfo {volume} {11}},\ \bibinfo {pages} {155} (\bibinfo
  {year} {2010})},\ \Eprint {http://arxiv.org/abs/1008.0098} {arXiv:1008.0098}
  \BibitemShut {NoStop}%
\bibitem [{\citenamefont {Del~Duca}\ \emph {et~al.}(2012)\citenamefont
  {Del~Duca}, \citenamefont {Duhr}, \citenamefont {Gardi}, \citenamefont
  {Magnea},\ and\ \citenamefont {White}}]{Bret:2011xm}%
  \BibitemOpen
  \bibfield  {author} {\bibinfo {author} {\bibfnamefont {V.}~\bibnamefont
  {Del~Duca}}, \bibinfo {author} {\bibfnamefont {C.}~\bibnamefont {Duhr}},
  \bibinfo {author} {\bibfnamefont {E.}~\bibnamefont {Gardi}}, \bibinfo
  {author} {\bibfnamefont {L.}~\bibnamefont {Magnea}}, \ and\ \bibinfo {author}
  {\bibfnamefont {C.~D.}\ \bibnamefont {White}},\ }\href {\doibase
  10.1103/PhysRevD.85.071104} {\bibfield  {journal} {\bibinfo  {journal} {Phys.
  Rev.}\ }\textbf {\bibinfo {volume} {D85}},\ \bibinfo {pages} {071104}
  (\bibinfo {year} {2012})},\ \Eprint {http://arxiv.org/abs/1108.5947}
  {arXiv:1108.5947} \BibitemShut {NoStop}%
\bibitem [{\citenamefont {Del~Duca}\ \emph {et~al.}(2011)\citenamefont
  {Del~Duca}, \citenamefont {Duhr}, \citenamefont {Gardi}, \citenamefont
  {Magnea},\ and\ \citenamefont {White}}]{DelDuca:2011ae}%
  \BibitemOpen
  \bibfield  {author} {\bibinfo {author} {\bibfnamefont {V.}~\bibnamefont
  {Del~Duca}}, \bibinfo {author} {\bibfnamefont {C.}~\bibnamefont {Duhr}},
  \bibinfo {author} {\bibfnamefont {E.}~\bibnamefont {Gardi}}, \bibinfo
  {author} {\bibfnamefont {L.}~\bibnamefont {Magnea}}, \ and\ \bibinfo {author}
  {\bibfnamefont {C.~D.}\ \bibnamefont {White}},\ }\href {\doibase
  10.1007/JHEP12(2011)021} {\bibfield  {journal} {\bibinfo  {journal} {JHEP}\
  }\textbf {\bibinfo {volume} {12}},\ \bibinfo {pages} {021} (\bibinfo {year}
  {2011})},\ \Eprint {http://arxiv.org/abs/1109.3581} {arXiv:1109.3581}
  \BibitemShut {NoStop}%
\bibitem [{\citenamefont {Ahrens}\ \emph {et~al.}(2012)\citenamefont {Ahrens},
  \citenamefont {Neubert},\ and\ \citenamefont {Vernazza}}]{Ahrens:2012qz}%
  \BibitemOpen
  \bibfield  {author} {\bibinfo {author} {\bibfnamefont {V.}~\bibnamefont
  {Ahrens}}, \bibinfo {author} {\bibfnamefont {M.}~\bibnamefont {Neubert}}, \
  and\ \bibinfo {author} {\bibfnamefont {L.}~\bibnamefont {Vernazza}},\ }\href
  {\doibase 10.1007/JHEP09(2012)138} {\bibfield  {journal} {\bibinfo  {journal}
  {JHEP}\ }\textbf {\bibinfo {volume} {09}},\ \bibinfo {pages} {138} (\bibinfo
  {year} {2012})},\ \Eprint {http://arxiv.org/abs/1208.4847} {arXiv:1208.4847}
  \BibitemShut {NoStop}%
\bibitem [{\citenamefont {Dukes}\ \emph {et~al.}(2013)\citenamefont {Dukes},
  \citenamefont {Gardi}, \citenamefont {Steingrimsson},\ and\ \citenamefont
  {White}}]{Dukes:2013wa}%
  \BibitemOpen
  \bibfield  {author} {\bibinfo {author} {\bibfnamefont {M.}~\bibnamefont
  {Dukes}}, \bibinfo {author} {\bibfnamefont {E.}~\bibnamefont {Gardi}},
  \bibinfo {author} {\bibfnamefont {E.}~\bibnamefont {Steingrimsson}}, \ and\
  \bibinfo {author} {\bibfnamefont {C.~D.}\ \bibnamefont {White}},\ }\href
  {\doibase 10.1016/j.jcta.2013.02.001} {\bibfield  {journal} {\bibinfo
  {journal} {J. Comb. Theory Ser.}\ }\textbf {\bibinfo {volume} {A120}},\
  \bibinfo {pages} {1012} (\bibinfo {year} {2013})},\ \Eprint
  {http://arxiv.org/abs/1301.6576} {arXiv:1301.6576} \BibitemShut {NoStop}%
\bibitem [{\citenamefont {Gardi}\ \emph {et~al.}(2013)\citenamefont {Gardi},
  \citenamefont {Smillie},\ and\ \citenamefont {White}}]{Gardi:2013ita}%
  \BibitemOpen
  \bibfield  {author} {\bibinfo {author} {\bibfnamefont {E.}~\bibnamefont
  {Gardi}}, \bibinfo {author} {\bibfnamefont {J.~M.}\ \bibnamefont {Smillie}},
  \ and\ \bibinfo {author} {\bibfnamefont {C.~D.}\ \bibnamefont {White}},\
  }\href {\doibase 10.1007/JHEP06(2013)088} {\bibfield  {journal} {\bibinfo
  {journal} {JHEP}\ }\textbf {\bibinfo {volume} {06}},\ \bibinfo {pages} {088}
  (\bibinfo {year} {2013})},\ \Eprint {http://arxiv.org/abs/1304.7040}
  {arXiv:1304.7040} \BibitemShut {NoStop}%
\bibitem [{\citenamefont {Caron-Huot}(2015)}]{Caron-Huot:2013fea}%
  \BibitemOpen
  \bibfield  {author} {\bibinfo {author} {\bibfnamefont {S.}~\bibnamefont
  {Caron-Huot}},\ }\href {\doibase 10.1007/JHEP05(2015)093} {\bibfield
  {journal} {\bibinfo  {journal} {JHEP}\ }\textbf {\bibinfo {volume} {05}},\
  \bibinfo {pages} {093} (\bibinfo {year} {2015})},\ \Eprint
  {http://arxiv.org/abs/1309.6521} {arXiv:1309.6521} \BibitemShut {NoStop}%
\bibitem [{\citenamefont {Falcioni}\ \emph {et~al.}(2014)\citenamefont
  {Falcioni}, \citenamefont {Gardi}, \citenamefont {Harley}, \citenamefont
  {Magnea},\ and\ \citenamefont {White}}]{Falcioni:2014pka}%
  \BibitemOpen
  \bibfield  {author} {\bibinfo {author} {\bibfnamefont {G.}~\bibnamefont
  {Falcioni}}, \bibinfo {author} {\bibfnamefont {E.}~\bibnamefont {Gardi}},
  \bibinfo {author} {\bibfnamefont {M.}~\bibnamefont {Harley}}, \bibinfo
  {author} {\bibfnamefont {L.}~\bibnamefont {Magnea}}, \ and\ \bibinfo {author}
  {\bibfnamefont {C.~D.}\ \bibnamefont {White}},\ }\href {\doibase
  10.1007/JHEP10(2014)010} {\bibfield  {journal} {\bibinfo  {journal} {JHEP}\
  }\textbf {\bibinfo {volume} {10}},\ \bibinfo {pages} {10} (\bibinfo {year}
  {2014})},\ \Eprint {http://arxiv.org/abs/1407.3477} {arXiv:1407.3477}
  \BibitemShut {NoStop}%
\bibitem [{\citenamefont {Boels}\ \emph {et~al.}(2017)\citenamefont {Boels},
  \citenamefont {Huber},\ and\ \citenamefont {Yang}}]{Boels:2017skl}%
  \BibitemOpen
  \bibfield  {author} {\bibinfo {author} {\bibfnamefont {R.~H.}\ \bibnamefont
  {Boels}}, \bibinfo {author} {\bibfnamefont {T.}~\bibnamefont {Huber}}, \ and\
  \bibinfo {author} {\bibfnamefont {G.}~\bibnamefont {Yang}},\ }\href {\doibase
  10.1103/PhysRevLett.119.201601} {\bibfield  {journal} {\bibinfo  {journal}
  {Phys. Rev. Lett.}\ }\textbf {\bibinfo {volume} {119}},\ \bibinfo {pages}
  {201601} (\bibinfo {year} {2017})},\ \Eprint
  {http://arxiv.org/abs/1705.03444} {arXiv:1705.03444} \BibitemShut {NoStop}%
\bibitem [{\citenamefont {Almelid}\ \emph {et~al.}(2016)\citenamefont
  {Almelid}, \citenamefont {Duhr},\ and\ \citenamefont
  {Gardi}}]{Almelid:2015jia}%
  \BibitemOpen
  \bibfield  {author} {\bibinfo {author} {\bibfnamefont {{\O}.}~\bibnamefont
  {Almelid}}, \bibinfo {author} {\bibfnamefont {C.}~\bibnamefont {Duhr}}, \
  and\ \bibinfo {author} {\bibfnamefont {E.}~\bibnamefont {Gardi}},\ }\href
  {\doibase 10.1103/PhysRevLett.117.172002} {\bibfield  {journal} {\bibinfo
  {journal} {Phys. Rev. Lett.}\ }\textbf {\bibinfo {volume} {117}},\ \bibinfo
  {pages} {172002} (\bibinfo {year} {2016})},\ \Eprint
  {http://arxiv.org/abs/1507.00047} {arXiv:1507.00047} \BibitemShut {NoStop}%
\bibitem [{\citenamefont {Henn}\ and\ \citenamefont
  {Mistlberger}(2016)}]{Henn:2016jdu}%
  \BibitemOpen
  \bibfield  {author} {\bibinfo {author} {\bibfnamefont {J.~M.}\ \bibnamefont
  {Henn}}\ and\ \bibinfo {author} {\bibfnamefont {B.}~\bibnamefont
  {Mistlberger}},\ }\href {\doibase 10.1103/PhysRevLett.117.171601} {\bibfield
  {journal} {\bibinfo  {journal} {Phys. Rev. Lett.}\ }\textbf {\bibinfo
  {volume} {117}},\ \bibinfo {pages} {171601} (\bibinfo {year} {2016})},\
  \Eprint {http://arxiv.org/abs/1608.00850} {arXiv:1608.00850} \BibitemShut
  {NoStop}%
\bibitem [{\citenamefont {Caron-Huot}\ \emph {et~al.}(2018)\citenamefont
  {Caron-Huot}, \citenamefont {Gardi}, \citenamefont {Reichel},\ and\
  \citenamefont {Vernazza}}]{Caron-Huot:2017zfo}%
  \BibitemOpen
  \bibfield  {author} {\bibinfo {author} {\bibfnamefont {S.}~\bibnamefont
  {Caron-Huot}}, \bibinfo {author} {\bibfnamefont {E.}~\bibnamefont {Gardi}},
  \bibinfo {author} {\bibfnamefont {J.}~\bibnamefont {Reichel}}, \ and\
  \bibinfo {author} {\bibfnamefont {L.}~\bibnamefont {Vernazza}},\ }\href
  {\doibase 10.1007/JHEP03(2018)098} {\bibfield  {journal} {\bibinfo  {journal}
  {JHEP}\ }\textbf {\bibinfo {volume} {03}},\ \bibinfo {pages} {098} (\bibinfo
  {year} {2018})},\ \Eprint {http://arxiv.org/abs/1711.04850}
  {arXiv:1711.04850} \BibitemShut {NoStop}%
\bibitem [{\citenamefont {Almelid}\ \emph {et~al.}(2017)\citenamefont
  {Almelid}, \citenamefont {Duhr}, \citenamefont {Gardi}, \citenamefont
  {McLeod},\ and\ \citenamefont {White}}]{Almelid:2017qju}%
  \BibitemOpen
  \bibfield  {author} {\bibinfo {author} {\bibfnamefont {{\O}.}~\bibnamefont
  {Almelid}}, \bibinfo {author} {\bibfnamefont {C.}~\bibnamefont {Duhr}},
  \bibinfo {author} {\bibfnamefont {E.}~\bibnamefont {Gardi}}, \bibinfo
  {author} {\bibfnamefont {A.}~\bibnamefont {McLeod}}, \ and\ \bibinfo {author}
  {\bibfnamefont {C.~D.}\ \bibnamefont {White}},\ }\href {\doibase
  10.1007/JHEP09(2017)073} {\bibfield  {journal} {\bibinfo  {journal} {JHEP}\
  }\textbf {\bibinfo {volume} {09}},\ \bibinfo {pages} {073} (\bibinfo {year}
  {2017})},\ \Eprint {http://arxiv.org/abs/1706.10162} {arXiv:1706.10162}
  \BibitemShut {NoStop}%
\bibitem [{\citenamefont {Moch}\ \emph {et~al.}(2017)\citenamefont {Moch},
  \citenamefont {Ruijl}, \citenamefont {Ueda}, \citenamefont {Vermaseren},\
  and\ \citenamefont {Vogt}}]{Moch:2017uml}%
  \BibitemOpen
  \bibfield  {author} {\bibinfo {author} {\bibfnamefont {S.}~\bibnamefont
  {Moch}}, \bibinfo {author} {\bibfnamefont {B.}~\bibnamefont {Ruijl}},
  \bibinfo {author} {\bibfnamefont {T.}~\bibnamefont {Ueda}}, \bibinfo {author}
  {\bibfnamefont {J.~A.~M.}\ \bibnamefont {Vermaseren}}, \ and\ \bibinfo
  {author} {\bibfnamefont {A.}~\bibnamefont {Vogt}},\ }\href {\doibase
  10.1007/JHEP10(2017)041} {\bibfield  {journal} {\bibinfo  {journal} {JHEP}\
  }\textbf {\bibinfo {volume} {10}},\ \bibinfo {pages} {041} (\bibinfo {year}
  {2017})},\ \Eprint {http://arxiv.org/abs/1707.08315} {arXiv:1707.08315}
  \BibitemShut {NoStop}%
\bibitem [{\citenamefont {Grozin}\ \emph {et~al.}(2017)\citenamefont {Grozin},
  \citenamefont {Henn},\ and\ \citenamefont {Stahlhofen}}]{Grozin:2017css}%
  \BibitemOpen
  \bibfield  {author} {\bibinfo {author} {\bibfnamefont {A.}~\bibnamefont
  {Grozin}}, \bibinfo {author} {\bibfnamefont {J.}~\bibnamefont {Henn}}, \ and\
  \bibinfo {author} {\bibfnamefont {M.}~\bibnamefont {Stahlhofen}},\ }\href
  {\doibase 10.1007/JHEP10(2017)052} {\bibfield  {journal} {\bibinfo  {journal}
  {JHEP}\ }\textbf {\bibinfo {volume} {10}},\ \bibinfo {pages} {052} (\bibinfo
  {year} {2017})},\ \Eprint {http://arxiv.org/abs/1708.01221}
  {arXiv:1708.01221} \BibitemShut {NoStop}%
\bibitem [{\citenamefont {Boels}\ \emph {et~al.}(2018)\citenamefont {Boels},
  \citenamefont {Huber},\ and\ \citenamefont {Yang}}]{Boels:2017ftb}%
  \BibitemOpen
  \bibfield  {author} {\bibinfo {author} {\bibfnamefont {R.~H.}\ \bibnamefont
  {Boels}}, \bibinfo {author} {\bibfnamefont {T.}~\bibnamefont {Huber}}, \ and\
  \bibinfo {author} {\bibfnamefont {G.}~\bibnamefont {Yang}},\ }\href {\doibase
  10.1007/JHEP01(2018)153} {\bibfield  {journal} {\bibinfo  {journal} {JHEP}\
  }\textbf {\bibinfo {volume} {01}},\ \bibinfo {pages} {153} (\bibinfo {year}
  {2018})},\ \Eprint {http://arxiv.org/abs/1711.08449} {arXiv:1711.08449}
  \BibitemShut {NoStop}%
\bibitem [{\citenamefont {von Manteuffel}\ \emph {et~al.}(2015)\citenamefont
  {von Manteuffel}, \citenamefont {Panzer},\ and\ \citenamefont
  {Schabinger}}]{vonManteuffel:2014qoa}%
  \BibitemOpen
  \bibfield  {author} {\bibinfo {author} {\bibfnamefont {A.}~\bibnamefont {von
  Manteuffel}}, \bibinfo {author} {\bibfnamefont {E.}~\bibnamefont {Panzer}}, \
  and\ \bibinfo {author} {\bibfnamefont {R.~M.}\ \bibnamefont {Schabinger}},\
  }\href {\doibase 10.1007/JHEP02(2015)120} {\bibfield  {journal} {\bibinfo
  {journal} {JHEP}\ }\textbf {\bibinfo {volume} {02}},\ \bibinfo {pages} {120}
  (\bibinfo {year} {2015})},\ \Eprint {http://arxiv.org/abs/1411.7392}
  {arXiv:1411.7392} \BibitemShut {NoStop}%
\bibitem [{\citenamefont {von Manteuffel}\ \emph {et~al.}(2016)\citenamefont
  {von Manteuffel}, \citenamefont {Panzer},\ and\ \citenamefont
  {Schabinger}}]{vonManteuffel:2015gxa}%
  \BibitemOpen
  \bibfield  {author} {\bibinfo {author} {\bibfnamefont {A.}~\bibnamefont {von
  Manteuffel}}, \bibinfo {author} {\bibfnamefont {E.}~\bibnamefont {Panzer}}, \
  and\ \bibinfo {author} {\bibfnamefont {R.~M.}\ \bibnamefont {Schabinger}},\
  }\href {\doibase 10.1103/PhysRevD.93.125014} {\bibfield  {journal} {\bibinfo
  {journal} {Phys. Rev.}\ }\textbf {\bibinfo {volume} {D93}},\ \bibinfo {pages}
  {125014} (\bibinfo {year} {2016})},\ \Eprint
  {http://arxiv.org/abs/1510.06758} {arXiv:1510.06758} \BibitemShut {NoStop}%
\bibitem [{\citenamefont {Kotikov}(2010)}]{Kotikov:2010gf}%
  \BibitemOpen
  \bibfield  {author} {\bibinfo {author} {\bibfnamefont {A.~V.}\ \bibnamefont
  {Kotikov}},\ }in\ \href
  {http://inspirehep.net/record/856436/files/arXiv:1005.5029.pdf} {\emph
  {\bibinfo {booktitle} {{Diakonov, D. (ed.): Subtleties in quantum field
  theory}}}}\ (\bibinfo {year} {2010})\ pp.\ \bibinfo {pages} {150--174},\
  \Eprint {http://arxiv.org/abs/1005.5029} {arXiv:1005.5029} \BibitemShut
  {NoStop}%
\bibitem [{\citenamefont {Henn}(2013)}]{Henn:2013pwa}%
  \BibitemOpen
  \bibfield  {author} {\bibinfo {author} {\bibfnamefont {J.~M.}\ \bibnamefont
  {Henn}},\ }\href {\doibase 10.1103/PhysRevLett.110.251601} {\bibfield
  {journal} {\bibinfo  {journal} {Phys. Rev. Lett.}\ }\textbf {\bibinfo
  {volume} {110}},\ \bibinfo {pages} {251601} (\bibinfo {year} {2013})},\
  \Eprint {http://arxiv.org/abs/1304.1806} {arXiv:1304.1806} \BibitemShut
  {NoStop}%
\bibitem [{\citenamefont {Bonciani}\ \emph {et~al.}(2016)\citenamefont
  {Bonciani}, \citenamefont {Del~Duca}, \citenamefont {Frellesvig},
  \citenamefont {Henn}, \citenamefont {Moriello},\ and\ \citenamefont
  {Smirnov}}]{Bonciani:2016qxi}%
  \BibitemOpen
  \bibfield  {author} {\bibinfo {author} {\bibfnamefont {R.}~\bibnamefont
  {Bonciani}}, \bibinfo {author} {\bibfnamefont {V.}~\bibnamefont {Del~Duca}},
  \bibinfo {author} {\bibfnamefont {H.}~\bibnamefont {Frellesvig}}, \bibinfo
  {author} {\bibfnamefont {J.~M.}\ \bibnamefont {Henn}}, \bibinfo {author}
  {\bibfnamefont {F.}~\bibnamefont {Moriello}}, \ and\ \bibinfo {author}
  {\bibfnamefont {V.~A.}\ \bibnamefont {Smirnov}},\ }\href {\doibase
  10.1007/JHEP12(2016)096} {\bibfield  {journal} {\bibinfo  {journal} {JHEP}\
  }\textbf {\bibinfo {volume} {12}},\ \bibinfo {pages} {096} (\bibinfo {year}
  {2016})},\ \Eprint {http://arxiv.org/abs/1609.06685} {arXiv:1609.06685}
  \BibitemShut {NoStop}%
\bibitem [{\citenamefont {von Manteuffel}\ and\ \citenamefont
  {Tancredi}(2017)}]{vonManteuffel:2017hms}%
  \BibitemOpen
  \bibfield  {author} {\bibinfo {author} {\bibfnamefont {A.}~\bibnamefont {von
  Manteuffel}}\ and\ \bibinfo {author} {\bibfnamefont {L.}~\bibnamefont
  {Tancredi}},\ }\href {\doibase 10.1007/JHEP06(2017)127} {\bibfield  {journal}
  {\bibinfo  {journal} {JHEP}\ }\textbf {\bibinfo {volume} {06}},\ \bibinfo
  {pages} {127} (\bibinfo {year} {2017})},\ \Eprint
  {http://arxiv.org/abs/1701.05905} {arXiv:1701.05905} \BibitemShut {NoStop}%
\bibitem [{\citenamefont {Ablinger}\ \emph {et~al.}(2018)\citenamefont
  {Ablinger}, \citenamefont {Bl{\"u}mlein}, \citenamefont {De~Freitas},
  \citenamefont {van Hoeij}, \citenamefont {Imamoglu}, \citenamefont {Raab},
  \citenamefont {Radu},\ and\ \citenamefont {Schneider}}]{Ablinger:2017bjx}%
  \BibitemOpen
  \bibfield  {author} {\bibinfo {author} {\bibfnamefont {J.}~\bibnamefont
  {Ablinger}}, \bibinfo {author} {\bibfnamefont {J.}~\bibnamefont
  {Bl{\"u}mlein}}, \bibinfo {author} {\bibfnamefont {A.}~\bibnamefont
  {De~Freitas}}, \bibinfo {author} {\bibfnamefont {M.}~\bibnamefont {van
  Hoeij}}, \bibinfo {author} {\bibfnamefont {E.}~\bibnamefont {Imamoglu}},
  \bibinfo {author} {\bibfnamefont {C.~G.}\ \bibnamefont {Raab}}, \bibinfo
  {author} {\bibfnamefont {C.~S.}\ \bibnamefont {Radu}}, \ and\ \bibinfo
  {author} {\bibfnamefont {C.}~\bibnamefont {Schneider}},\ }\href {\doibase
  10.1063/1.4986417} {\bibfield  {journal} {\bibinfo  {journal} {J. Math.
  Phys.}\ }\textbf {\bibinfo {volume} {59}},\ \bibinfo {pages} {062305}
  (\bibinfo {year} {2018})},\ \Eprint {http://arxiv.org/abs/1706.01299}
  {arXiv:1706.01299 [hep-th]} \BibitemShut {NoStop}%
\bibitem [{\citenamefont {Remiddi}\ and\ \citenamefont
  {Tancredi}(2017)}]{Remiddi:2017har}%
  \BibitemOpen
  \bibfield  {author} {\bibinfo {author} {\bibfnamefont {E.}~\bibnamefont
  {Remiddi}}\ and\ \bibinfo {author} {\bibfnamefont {L.}~\bibnamefont
  {Tancredi}},\ }\href {\doibase 10.1016/j.nuclphysb.2017.10.007} {\bibfield
  {journal} {\bibinfo  {journal} {Nucl. Phys.}\ }\textbf {\bibinfo {volume}
  {B925}},\ \bibinfo {pages} {212} (\bibinfo {year} {2017})},\ \Eprint
  {http://arxiv.org/abs/1709.03622} {arXiv:1709.03622} \BibitemShut {NoStop}%
\bibitem [{\citenamefont {Bl{\"u}mlein}\ \emph {et~al.}(2017)\citenamefont
  {Bl{\"u}mlein}, \citenamefont {Ablinger}, \citenamefont {Behring},
  \citenamefont {De~Freitas}, \citenamefont {Imamoglu}, \citenamefont {van
  Hoeij}, \citenamefont {von Manteuffel}, \citenamefont {Raab}, \citenamefont
  {Radu},\ and\ \citenamefont {Schneider}}]{Bluemlein:2017ibj}%
  \BibitemOpen
  \bibfield  {author} {\bibinfo {author} {\bibfnamefont {J.}~\bibnamefont
  {Bl{\"u}mlein}}, \bibinfo {author} {\bibfnamefont {J.}~\bibnamefont
  {Ablinger}}, \bibinfo {author} {\bibfnamefont {A.}~\bibnamefont {Behring}},
  \bibinfo {author} {\bibfnamefont {A.}~\bibnamefont {De~Freitas}}, \bibinfo
  {author} {\bibfnamefont {E.}~\bibnamefont {Imamoglu}}, \bibinfo {author}
  {\bibfnamefont {M.}~\bibnamefont {van Hoeij}}, \bibinfo {author}
  {\bibfnamefont {A.}~\bibnamefont {von Manteuffel}}, \bibinfo {author}
  {\bibfnamefont {C.~G.}\ \bibnamefont {Raab}}, \bibinfo {author}
  {\bibfnamefont {C.~S.}\ \bibnamefont {Radu}}, \ and\ \bibinfo {author}
  {\bibfnamefont {C.}~\bibnamefont {Schneider}},\ }\bibfield  {booktitle}
  {\emph {\bibinfo {booktitle} {{Proceedings, 13th International Symposium on
  Radiative Corrections: Application of Quantum Field Theory to Phenomenology
  (RADCOR2017): St. Gilgen, Austria, September 24-29, 2017}}},\ }\href
  {\doibase 10.22323/1.290.0069} {\bibfield  {journal} {\bibinfo  {journal}
  {PoS}\ }\textbf {\bibinfo {volume} {RADCOR2017}},\ \bibinfo {pages} {069}
  (\bibinfo {year} {2017})},\ \Eprint {http://arxiv.org/abs/1711.09742}
  {arXiv:1711.09742 [hep-ph]} \BibitemShut {NoStop}%
\bibitem [{\citenamefont {Bourjaily}\ \emph {et~al.}(2018)\citenamefont
  {Bourjaily}, \citenamefont {McLeod}, \citenamefont {Spradlin}, \citenamefont
  {von Hippel},\ and\ \citenamefont {Wilhelm}}]{Bourjaily:2017bsb}%
  \BibitemOpen
  \bibfield  {author} {\bibinfo {author} {\bibfnamefont {J.~L.}\ \bibnamefont
  {Bourjaily}}, \bibinfo {author} {\bibfnamefont {A.~J.}\ \bibnamefont
  {McLeod}}, \bibinfo {author} {\bibfnamefont {M.}~\bibnamefont {Spradlin}},
  \bibinfo {author} {\bibfnamefont {M.}~\bibnamefont {von Hippel}}, \ and\
  \bibinfo {author} {\bibfnamefont {M.}~\bibnamefont {Wilhelm}},\ }\href
  {\doibase 10.1103/PhysRevLett.120.121603} {\bibfield  {journal} {\bibinfo
  {journal} {Phys. Rev. Lett.}\ }\textbf {\bibinfo {volume} {120}},\ \bibinfo
  {pages} {121603} (\bibinfo {year} {2018})},\ \Eprint
  {http://arxiv.org/abs/1712.02785} {arXiv:1712.02785} \BibitemShut {NoStop}%
\bibitem [{\citenamefont {Hidding}\ and\ \citenamefont
  {Moriello}(2017)}]{Hidding:2017jkk}%
  \BibitemOpen
  \bibfield  {author} {\bibinfo {author} {\bibfnamefont {M.}~\bibnamefont
  {Hidding}}\ and\ \bibinfo {author} {\bibfnamefont {F.}~\bibnamefont
  {Moriello}},\ }\href@noop {} {\  (\bibinfo {year} {2017})},\ \Eprint
  {http://arxiv.org/abs/1712.04441} {arXiv:1712.04441} \BibitemShut {NoStop}%
\bibitem [{\citenamefont {Br{\"o}del}\ \emph
  {et~al.}(2018{\natexlab{a}})\citenamefont {Br{\"o}del}, \citenamefont {Duhr},
  \citenamefont {Dulat},\ and\ \citenamefont {Tancredi}}]{Broedel:2017kkb}%
  \BibitemOpen
  \bibfield  {author} {\bibinfo {author} {\bibfnamefont {J.}~\bibnamefont
  {Br{\"o}del}}, \bibinfo {author} {\bibfnamefont {C.}~\bibnamefont {Duhr}},
  \bibinfo {author} {\bibfnamefont {F.}~\bibnamefont {Dulat}}, \ and\ \bibinfo
  {author} {\bibfnamefont {L.}~\bibnamefont {Tancredi}},\ }\href {\doibase
  10.1007/JHEP05(2018)093} {\bibfield  {journal} {\bibinfo  {journal} {JHEP}\
  }\textbf {\bibinfo {volume} {05}},\ \bibinfo {pages} {093} (\bibinfo {year}
  {2018}{\natexlab{a}})},\ \Eprint {http://arxiv.org/abs/1712.07089}
  {arXiv:1712.07089} \BibitemShut {NoStop}%
\bibitem [{\citenamefont {Br{\"o}del}\ \emph
  {et~al.}(2018{\natexlab{b}})\citenamefont {Br{\"o}del}, \citenamefont {Duhr},
  \citenamefont {Dulat},\ and\ \citenamefont {Tancredi}}]{Broedel:2017siw}%
  \BibitemOpen
  \bibfield  {author} {\bibinfo {author} {\bibfnamefont {J.}~\bibnamefont
  {Br{\"o}del}}, \bibinfo {author} {\bibfnamefont {C.}~\bibnamefont {Duhr}},
  \bibinfo {author} {\bibfnamefont {F.}~\bibnamefont {Dulat}}, \ and\ \bibinfo
  {author} {\bibfnamefont {L.}~\bibnamefont {Tancredi}},\ }\href {\doibase
  10.1103/PhysRevD.97.116009} {\bibfield  {journal} {\bibinfo  {journal} {Phys.
  Rev.}\ }\textbf {\bibinfo {volume} {D97}},\ \bibinfo {pages} {116009}
  (\bibinfo {year} {2018}{\natexlab{b}})},\ \Eprint
  {http://arxiv.org/abs/1712.07095} {arXiv:1712.07095 [hep-ph]} \BibitemShut
  {NoStop}%
\bibitem [{\citenamefont {Adams}\ and\ \citenamefont
  {Weinzierl}(2018)}]{Adams:2018yfj}%
  \BibitemOpen
  \bibfield  {author} {\bibinfo {author} {\bibfnamefont {L.}~\bibnamefont
  {Adams}}\ and\ \bibinfo {author} {\bibfnamefont {S.}~\bibnamefont
  {Weinzierl}},\ }\href {\doibase 10.1016/j.physletb.2018.04.002} {\bibfield
  {journal} {\bibinfo  {journal} {Phys. Lett.}\ }\textbf {\bibinfo {volume}
  {B781}},\ \bibinfo {pages} {270} (\bibinfo {year} {2018})},\ \Eprint
  {http://arxiv.org/abs/1802.05020} {arXiv:1802.05020} \BibitemShut {NoStop}%
\bibitem [{\citenamefont {Br{\"o}del}\ \emph
  {et~al.}(2018{\natexlab{c}})\citenamefont {Br{\"o}del}, \citenamefont {Duhr},
  \citenamefont {Dulat}, \citenamefont {Penante},\ and\ \citenamefont
  {Tancredi}}]{Broedel:2018iwv}%
  \BibitemOpen
  \bibfield  {author} {\bibinfo {author} {\bibfnamefont {J.}~\bibnamefont
  {Br{\"o}del}}, \bibinfo {author} {\bibfnamefont {C.}~\bibnamefont {Duhr}},
  \bibinfo {author} {\bibfnamefont {F.}~\bibnamefont {Dulat}}, \bibinfo
  {author} {\bibfnamefont {B.}~\bibnamefont {Penante}}, \ and\ \bibinfo
  {author} {\bibfnamefont {L.}~\bibnamefont {Tancredi}},\ }\href {\doibase
  10.1007/JHEP08(2018)014} {\bibfield  {journal} {\bibinfo  {journal} {JHEP}\
  }\textbf {\bibinfo {volume} {08}},\ \bibinfo {pages} {014} (\bibinfo {year}
  {2018}{\natexlab{c}})},\ \Eprint {http://arxiv.org/abs/1803.10256}
  {arXiv:1803.10256 [hep-th]} \BibitemShut {NoStop}%
\bibitem [{\citenamefont {Lee}(2018)}]{Lee:2018jsw}%
  \BibitemOpen
  \bibfield  {author} {\bibinfo {author} {\bibfnamefont {R.~N.}\ \bibnamefont
  {Lee}},\ }\href {\doibase 10.1007/JHEP10(2018)176} {\bibfield  {journal}
  {\bibinfo  {journal} {JHEP}\ }\textbf {\bibinfo {volume} {10}},\ \bibinfo
  {pages} {176} (\bibinfo {year} {2018})},\ \Eprint
  {http://arxiv.org/abs/1806.04846} {arXiv:1806.04846 [hep-ph]} \BibitemShut
  {NoStop}%
\bibitem [{\citenamefont {Br{\"o}del}\ \emph {et~al.}(2019)\citenamefont
  {Br{\"o}del}, \citenamefont {Duhr}, \citenamefont {Dulat}, \citenamefont
  {Penante},\ and\ \citenamefont {Tancredi}}]{Broedel:2018qkq}%
  \BibitemOpen
  \bibfield  {author} {\bibinfo {author} {\bibfnamefont {J.}~\bibnamefont
  {Br{\"o}del}}, \bibinfo {author} {\bibfnamefont {C.}~\bibnamefont {Duhr}},
  \bibinfo {author} {\bibfnamefont {F.}~\bibnamefont {Dulat}}, \bibinfo
  {author} {\bibfnamefont {B.}~\bibnamefont {Penante}}, \ and\ \bibinfo
  {author} {\bibfnamefont {L.}~\bibnamefont {Tancredi}},\ }\href {\doibase
  10.1007/JHEP01(2019)023} {\bibfield  {journal} {\bibinfo  {journal} {JHEP}\
  }\textbf {\bibinfo {volume} {01}},\ \bibinfo {pages} {023} (\bibinfo {year}
  {2019})},\ \Eprint {http://arxiv.org/abs/1809.10698} {arXiv:1809.10698
  [hep-th]} \BibitemShut {NoStop}%
\bibitem [{\citenamefont {Adams}\ \emph
  {et~al.}(2018{\natexlab{a}})\citenamefont {Adams}, \citenamefont {Chaubey},\
  and\ \citenamefont {Weinzierl}}]{Adams:2018bsn}%
  \BibitemOpen
  \bibfield  {author} {\bibinfo {author} {\bibfnamefont {L.}~\bibnamefont
  {Adams}}, \bibinfo {author} {\bibfnamefont {E.}~\bibnamefont {Chaubey}}, \
  and\ \bibinfo {author} {\bibfnamefont {S.}~\bibnamefont {Weinzierl}},\ }\href
  {\doibase 10.1103/PhysRevLett.121.142001} {\bibfield  {journal} {\bibinfo
  {journal} {Phys. Rev. Lett.}\ }\textbf {\bibinfo {volume} {121}},\ \bibinfo
  {pages} {142001} (\bibinfo {year} {2018}{\natexlab{a}})},\ \Eprint
  {http://arxiv.org/abs/1804.11144} {arXiv:1804.11144 [hep-ph]} \BibitemShut
  {NoStop}%
\bibitem [{\citenamefont {Adams}\ \emph
  {et~al.}(2018{\natexlab{b}})\citenamefont {Adams}, \citenamefont {Chaubey},\
  and\ \citenamefont {Weinzierl}}]{Adams:2018kez}%
  \BibitemOpen
  \bibfield  {author} {\bibinfo {author} {\bibfnamefont {L.}~\bibnamefont
  {Adams}}, \bibinfo {author} {\bibfnamefont {E.}~\bibnamefont {Chaubey}}, \
  and\ \bibinfo {author} {\bibfnamefont {S.}~\bibnamefont {Weinzierl}},\ }\href
  {\doibase 10.1007/JHEP10(2018)206} {\bibfield  {journal} {\bibinfo  {journal}
  {JHEP}\ }\textbf {\bibinfo {volume} {10}},\ \bibinfo {pages} {206} (\bibinfo
  {year} {2018}{\natexlab{b}})},\ \Eprint {http://arxiv.org/abs/1806.04981}
  {arXiv:1806.04981 [hep-ph]} \BibitemShut {NoStop}%
\bibitem [{\citenamefont {von Manteuffel}\ and\ \citenamefont
  {Schabinger}(2017)}]{vonManteuffel:2017myy}%
  \BibitemOpen
  \bibfield  {author} {\bibinfo {author} {\bibfnamefont {A.}~\bibnamefont {von
  Manteuffel}}\ and\ \bibinfo {author} {\bibfnamefont {R.~M.}\ \bibnamefont
  {Schabinger}},\ }\href {\doibase 10.1007/JHEP04(2017)129} {\bibfield
  {journal} {\bibinfo  {journal} {JHEP}\ }\textbf {\bibinfo {volume} {04}},\
  \bibinfo {pages} {129} (\bibinfo {year} {2017})},\ \Eprint
  {http://arxiv.org/abs/1701.06583} {arXiv:1701.06583} \BibitemShut {NoStop}%
\bibitem [{\citenamefont {Lee}(2010{\natexlab{a}})}]{Lee:2009dh}%
  \BibitemOpen
  \bibfield  {author} {\bibinfo {author} {\bibfnamefont {R.~N.}\ \bibnamefont
  {Lee}},\ }\href {\doibase 10.1016/j.nuclphysb.2009.12.025} {\bibfield
  {journal} {\bibinfo  {journal} {Nucl. Phys.}\ }\textbf {\bibinfo {volume}
  {B830}},\ \bibinfo {pages} {474} (\bibinfo {year} {2010}{\natexlab{a}})},\
  \Eprint {http://arxiv.org/abs/0911.0252} {arXiv:0911.0252} \BibitemShut
  {NoStop}%
\bibitem [{\citenamefont {Henn}\ \emph {et~al.}(2013)\citenamefont {Henn},
  \citenamefont {Smirnov},\ and\ \citenamefont {Smirnov}}]{Henn:2013fah}%
  \BibitemOpen
  \bibfield  {author} {\bibinfo {author} {\bibfnamefont {J.~M.}\ \bibnamefont
  {Henn}}, \bibinfo {author} {\bibfnamefont {A.~V.}\ \bibnamefont {Smirnov}}, \
  and\ \bibinfo {author} {\bibfnamefont {V.~A.}\ \bibnamefont {Smirnov}},\
  }\href {\doibase 10.1007/JHEP07(2013)128} {\bibfield  {journal} {\bibinfo
  {journal} {JHEP}\ }\textbf {\bibinfo {volume} {07}},\ \bibinfo {pages} {128}
  (\bibinfo {year} {2013})},\ \Eprint {http://arxiv.org/abs/1306.2799}
  {arXiv:1306.2799} \BibitemShut {NoStop}%
\bibitem [{\citenamefont {von Manteuffel}\ and\ \citenamefont
  {Studerus}()}]{vonManteuffel:2012np}%
  \BibitemOpen
  \bibfield  {author} {\bibinfo {author} {\bibfnamefont {A.}~\bibnamefont {von
  Manteuffel}}\ and\ \bibinfo {author} {\bibfnamefont {C.}~\bibnamefont
  {Studerus}},\ }\href@noop {} {\ }\Eprint {http://arxiv.org/abs/1201.4330}
  {arXiv:1201.4330} \BibitemShut {NoStop}%
\bibitem [{\citenamefont {Studerus}(2010)}]{Studerus:2009ye}%
  \BibitemOpen
  \bibfield  {author} {\bibinfo {author} {\bibfnamefont {C.}~\bibnamefont
  {Studerus}},\ }\href {\doibase 10.1016/j.cpc.2010.03.012} {\bibfield
  {journal} {\bibinfo  {journal} {Comput. Phys. Commun.}\ }\textbf {\bibinfo
  {volume} {181}},\ \bibinfo {pages} {1293} (\bibinfo {year} {2010})},\ \Eprint
  {http://arxiv.org/abs/0912.2546} {arXiv:0912.2546} \BibitemShut {NoStop}%
\bibitem [{\citenamefont {Bauer}\ \emph {et~al.}(2002)\citenamefont {Bauer},
  \citenamefont {Frink},\ and\ \citenamefont {Kreckel}}]{Bauer:2000cp}%
  \BibitemOpen
  \bibfield  {author} {\bibinfo {author} {\bibfnamefont {C.~W.}\ \bibnamefont
  {Bauer}}, \bibinfo {author} {\bibfnamefont {A.}~\bibnamefont {Frink}}, \ and\
  \bibinfo {author} {\bibfnamefont {R.}~\bibnamefont {Kreckel}},\ }\href@noop
  {} {\bibfield  {journal} {\bibinfo  {journal} {J. Symb. Comput.}\ }\textbf
  {\bibinfo {volume} {33}},\ \bibinfo {pages} {1} (\bibinfo {year} {2002})},\
  \Eprint {http://arxiv.org/abs/cs/0004015} {arXiv:cs/0004015} \BibitemShut
  {NoStop}%
\bibitem [{\citenamefont {Lewis}()}]{fermat}%
  \BibitemOpen
  \bibfield  {author} {\bibinfo {author} {\bibfnamefont {R.~H.}\ \bibnamefont
  {Lewis}},\ }\href@noop {} {\enquote {\bibinfo {title} {{\it Computer Algebra
  System {\tt Fermat}}},}\ }\bibinfo {howpublished}
  {\url{http://www.bway.net/∼lewis}}\BibitemShut {NoStop}%
\bibitem [{\citenamefont {van Neerven}(1986)}]{vanNeerven:1985xr}%
  \BibitemOpen
  \bibfield  {author} {\bibinfo {author} {\bibfnamefont {W.~L.}\ \bibnamefont
  {van Neerven}},\ }\href {\doibase 10.1016/0550-3213(86)90165-3} {\bibfield
  {journal} {\bibinfo  {journal} {Nucl. Phys.}\ }\textbf {\bibinfo {volume}
  {B268}},\ \bibinfo {pages} {453} (\bibinfo {year} {1986})}\BibitemShut
  {NoStop}%
\bibitem [{\citenamefont {Tarasov}(1996)}]{Tarasov:1996br}%
  \BibitemOpen
  \bibfield  {author} {\bibinfo {author} {\bibfnamefont {O.~V.}\ \bibnamefont
  {Tarasov}},\ }\href {\doibase 10.1103/PhysRevD.54.6479} {\bibfield  {journal}
  {\bibinfo  {journal} {Phys. Rev.}\ }\textbf {\bibinfo {volume} {D54}},\
  \bibinfo {pages} {6479} (\bibinfo {year} {1996})},\ \Eprint
  {http://arxiv.org/abs/hep-th/9606018} {arXiv:hep-th/9606018} \BibitemShut
  {NoStop}%
\bibitem [{\citenamefont {Tkachov}(1981)}]{Tkachov:1981wb}%
  \BibitemOpen
  \bibfield  {author} {\bibinfo {author} {\bibfnamefont {F.}~\bibnamefont
  {Tkachov}},\ }\href {\doibase 10.1016/0370-2693(81)90288-4} {\bibfield
  {journal} {\bibinfo  {journal} {Phys. Lett.}\ }\textbf {\bibinfo {volume}
  {B100}},\ \bibinfo {pages} {65} (\bibinfo {year} {1981})}\BibitemShut
  {NoStop}%
\bibitem [{\citenamefont {Chetyrkin}\ and\ \citenamefont
  {Tkachov}(1981)}]{Chetyrkin:1981qh}%
  \BibitemOpen
  \bibfield  {author} {\bibinfo {author} {\bibfnamefont {K.}~\bibnamefont
  {Chetyrkin}}\ and\ \bibinfo {author} {\bibfnamefont {F.}~\bibnamefont
  {Tkachov}},\ }\href {\doibase 10.1016/0550-3213(81)90199-1} {\bibfield
  {journal} {\bibinfo  {journal} {Nucl. Phys.}\ }\textbf {\bibinfo {volume}
  {B192}},\ \bibinfo {pages} {159} (\bibinfo {year} {1981})}\BibitemShut
  {NoStop}%
\bibitem [{\citenamefont {Bern}\ \emph
  {et~al.}(1994{\natexlab{a}})\citenamefont {Bern}, \citenamefont {Dixon},\
  and\ \citenamefont {Kosower}}]{Bern:1993kr}%
  \BibitemOpen
  \bibfield  {author} {\bibinfo {author} {\bibfnamefont {Z.}~\bibnamefont
  {Bern}}, \bibinfo {author} {\bibfnamefont {L.~J.}\ \bibnamefont {Dixon}}, \
  and\ \bibinfo {author} {\bibfnamefont {D.~A.}\ \bibnamefont {Kosower}},\
  }\href {\doibase 10.1016/0550-3213(94)90398-0} {\bibfield  {journal}
  {\bibinfo  {journal} {Nucl. Phys.}\ }\textbf {\bibinfo {volume} {B412}},\
  \bibinfo {pages} {751} (\bibinfo {year} {1994}{\natexlab{a}})},\ \Eprint
  {http://arxiv.org/abs/hep-ph/9306240} {arXiv:hep-ph/9306240} \BibitemShut
  {NoStop}%
\bibitem [{\citenamefont {Bern}\ \emph
  {et~al.}(1994{\natexlab{b}})\citenamefont {Bern}, \citenamefont {Dixon},
  \citenamefont {Dunbar},\ and\ \citenamefont {Kosower}}]{Bern:1994zx}%
  \BibitemOpen
  \bibfield  {author} {\bibinfo {author} {\bibfnamefont {Z.}~\bibnamefont
  {Bern}}, \bibinfo {author} {\bibfnamefont {L.~J.}\ \bibnamefont {Dixon}},
  \bibinfo {author} {\bibfnamefont {D.~C.}\ \bibnamefont {Dunbar}}, \ and\
  \bibinfo {author} {\bibfnamefont {D.~A.}\ \bibnamefont {Kosower}},\ }\href
  {\doibase 10.1016/0550-3213(94)90179-1} {\bibfield  {journal} {\bibinfo
  {journal} {Nucl. Phys.}\ }\textbf {\bibinfo {volume} {B425}},\ \bibinfo
  {pages} {217} (\bibinfo {year} {1994}{\natexlab{b}})},\ \Eprint
  {http://arxiv.org/abs/hep-ph/9403226} {arXiv:hep-ph/9403226} \BibitemShut
  {NoStop}%
\bibitem [{\citenamefont {Moch}\ \emph
  {et~al.}(2005{\natexlab{a}})\citenamefont {Moch}, \citenamefont
  {Vermaseren},\ and\ \citenamefont {Vogt}}]{Moch:2005id}%
  \BibitemOpen
  \bibfield  {author} {\bibinfo {author} {\bibfnamefont {S.}~\bibnamefont
  {Moch}}, \bibinfo {author} {\bibfnamefont {J.~A.~M.}\ \bibnamefont
  {Vermaseren}}, \ and\ \bibinfo {author} {\bibfnamefont {A.}~\bibnamefont
  {Vogt}},\ }\href {\doibase 10.1088/1126-6708/2005/08/049} {\bibfield
  {journal} {\bibinfo  {journal} {JHEP}\ }\textbf {\bibinfo {volume} {08}},\
  \bibinfo {pages} {049} (\bibinfo {year} {2005}{\natexlab{a}})},\ \Eprint
  {http://arxiv.org/abs/hep-ph/0507039} {arXiv:hep-ph/0507039} \BibitemShut
  {NoStop}%
\bibitem [{\citenamefont {Laporta}(2000)}]{Laporta:2001dd}%
  \BibitemOpen
  \bibfield  {author} {\bibinfo {author} {\bibfnamefont {S.}~\bibnamefont
  {Laporta}},\ }\href {\doibase 10.1016/S0217-751X(00)00215-7} {\bibfield
  {journal} {\bibinfo  {journal} {Int. J. Mod. Phys.}\ }\textbf {\bibinfo
  {volume} {A15}},\ \bibinfo {pages} {5087} (\bibinfo {year} {2000})},\ \Eprint
  {http://arxiv.org/abs/hep-ph/0102033} {arXiv:hep-ph/0102033} \BibitemShut
  {NoStop}%
\bibitem [{\citenamefont {Lee}(2010{\natexlab{b}})}]{Lee:2010wea}%
  \BibitemOpen
  \bibfield  {author} {\bibinfo {author} {\bibfnamefont {R.~N.}\ \bibnamefont
  {Lee}},\ }\href {\doibase 10.1016/j.nuclphysbps.2010.08.032} {\bibfield
  {journal} {\bibinfo  {journal} {Nucl. Phys. Proc. Suppl.}\ }\textbf {\bibinfo
  {volume} {205-206}},\ \bibinfo {pages} {135} (\bibinfo {year}
  {2010}{\natexlab{b}})},\ \Eprint {http://arxiv.org/abs/1007.2256}
  {arXiv:1007.2256} \BibitemShut {NoStop}%
\bibitem [{\citenamefont {Gehrmann}\ \emph {et~al.}(2005)\citenamefont
  {Gehrmann}, \citenamefont {Huber},\ and\ \citenamefont
  {Ma{\^i}tre}}]{Gehrmann:2005pd}%
  \BibitemOpen
  \bibfield  {author} {\bibinfo {author} {\bibfnamefont {T.}~\bibnamefont
  {Gehrmann}}, \bibinfo {author} {\bibfnamefont {T.}~\bibnamefont {Huber}}, \
  and\ \bibinfo {author} {\bibfnamefont {D.}~\bibnamefont {Ma{\^i}tre}},\
  }\href {\doibase 10.1016/j.physletb.2005.07.019} {\bibfield  {journal}
  {\bibinfo  {journal} {Phys. Lett.}\ }\textbf {\bibinfo {volume} {B622}},\
  \bibinfo {pages} {295} (\bibinfo {year} {2005})},\ \Eprint
  {http://arxiv.org/abs/hep-ph/0507061} {arXiv:hep-ph/0507061} \BibitemShut
  {NoStop}%
\bibitem [{\citenamefont {Gehrmann-De~Ridder}\ \emph
  {et~al.}(2004)\citenamefont {Gehrmann-De~Ridder}, \citenamefont {Gehrmann},\
  and\ \citenamefont {Heinrich}}]{Gehrmann-DeRidder:2003pne}%
  \BibitemOpen
  \bibfield  {author} {\bibinfo {author} {\bibfnamefont {A.}~\bibnamefont
  {Gehrmann-De~Ridder}}, \bibinfo {author} {\bibfnamefont {T.}~\bibnamefont
  {Gehrmann}}, \ and\ \bibinfo {author} {\bibfnamefont {G.}~\bibnamefont
  {Heinrich}},\ }\href {\doibase 10.1016/j.nuclphysb.2004.01.023} {\bibfield
  {journal} {\bibinfo  {journal} {Nucl. Phys.}\ }\textbf {\bibinfo {volume}
  {B682}},\ \bibinfo {pages} {265} (\bibinfo {year} {2004})},\ \Eprint
  {http://arxiv.org/abs/hep-ph/0311276} {arXiv:hep-ph/0311276} \BibitemShut
  {NoStop}%
\bibitem [{\citenamefont {Lee}\ and\ \citenamefont
  {Smirnov}(2012)}]{Lee:2012te}%
  \BibitemOpen
  \bibfield  {author} {\bibinfo {author} {\bibfnamefont {R.~N.}\ \bibnamefont
  {Lee}}\ and\ \bibinfo {author} {\bibfnamefont {V.~A.}\ \bibnamefont
  {Smirnov}},\ }\href {\doibase 10.1007/JHEP12(2012)104} {\bibfield  {journal}
  {\bibinfo  {journal} {JHEP}\ }\textbf {\bibinfo {volume} {12}},\ \bibinfo
  {pages} {104} (\bibinfo {year} {2012})},\ \Eprint
  {http://arxiv.org/abs/1209.0339} {arXiv:1209.0339} \BibitemShut {NoStop}%
\bibitem [{\citenamefont {Anastasiou}\ \emph {et~al.}(2013)\citenamefont
  {Anastasiou}, \citenamefont {Duhr}, \citenamefont {Dulat},\ and\
  \citenamefont {Mistlberger}}]{Anastasiou:2013srw}%
  \BibitemOpen
  \bibfield  {author} {\bibinfo {author} {\bibfnamefont {C.}~\bibnamefont
  {Anastasiou}}, \bibinfo {author} {\bibfnamefont {C.}~\bibnamefont {Duhr}},
  \bibinfo {author} {\bibfnamefont {F.}~\bibnamefont {Dulat}}, \ and\ \bibinfo
  {author} {\bibfnamefont {B.}~\bibnamefont {Mistlberger}},\ }\href {\doibase
  10.1007/JHEP07(2013)003} {\bibfield  {journal} {\bibinfo  {journal} {JHEP}\
  }\textbf {\bibinfo {volume} {07}},\ \bibinfo {pages} {003} (\bibinfo {year}
  {2013})},\ \Eprint {http://arxiv.org/abs/1302.4379} {arXiv:1302.4379}
  \BibitemShut {NoStop}%
\bibitem [{\citenamefont {Abreu}\ \emph {et~al.}(2014)\citenamefont {Abreu},
  \citenamefont {Britto}, \citenamefont {Duhr},\ and\ \citenamefont
  {Gardi}}]{Abreu:2014cla}%
  \BibitemOpen
  \bibfield  {author} {\bibinfo {author} {\bibfnamefont {S.}~\bibnamefont
  {Abreu}}, \bibinfo {author} {\bibfnamefont {R.}~\bibnamefont {Britto}},
  \bibinfo {author} {\bibfnamefont {C.}~\bibnamefont {Duhr}}, \ and\ \bibinfo
  {author} {\bibfnamefont {E.}~\bibnamefont {Gardi}},\ }\href {\doibase
  10.1007/JHEP10(2014)125} {\bibfield  {journal} {\bibinfo  {journal} {JHEP}\
  }\textbf {\bibinfo {volume} {10}},\ \bibinfo {pages} {125} (\bibinfo {year}
  {2014})},\ \Eprint {http://arxiv.org/abs/1401.3546} {arXiv:1401.3546}
  \BibitemShut {NoStop}%
\bibitem [{\citenamefont {Primo}\ and\ \citenamefont
  {Tancredi}(2017{\natexlab{a}})}]{Primo:2016ebd}%
  \BibitemOpen
  \bibfield  {author} {\bibinfo {author} {\bibfnamefont {A.}~\bibnamefont
  {Primo}}\ and\ \bibinfo {author} {\bibfnamefont {L.}~\bibnamefont
  {Tancredi}},\ }\href {\doibase 10.1016/j.nuclphysb.2016.12.021} {\bibfield
  {journal} {\bibinfo  {journal} {Nucl. Phys.}\ }\textbf {\bibinfo {volume}
  {B916}},\ \bibinfo {pages} {94} (\bibinfo {year} {2017}{\natexlab{a}})},\
  \Eprint {http://arxiv.org/abs/1610.08397} {arXiv:1610.08397} \BibitemShut
  {NoStop}%
\bibitem [{\citenamefont {Frellesvig}\ and\ \citenamefont
  {Papadopoulos}(2017)}]{Frellesvig:2017aai}%
  \BibitemOpen
  \bibfield  {author} {\bibinfo {author} {\bibfnamefont {H.}~\bibnamefont
  {Frellesvig}}\ and\ \bibinfo {author} {\bibfnamefont {C.~G.}\ \bibnamefont
  {Papadopoulos}},\ }\href {\doibase 10.1007/JHEP04(2017)083} {\bibfield
  {journal} {\bibinfo  {journal} {JHEP}\ }\textbf {\bibinfo {volume} {04}},\
  \bibinfo {pages} {083} (\bibinfo {year} {2017})},\ \Eprint
  {http://arxiv.org/abs/1701.07356} {arXiv:1701.07356} \BibitemShut {NoStop}%
\bibitem [{\citenamefont {Bosma}\ \emph {et~al.}(2017)\citenamefont {Bosma},
  \citenamefont {Sogaard},\ and\ \citenamefont {Zhang}}]{Bosma:2017ens}%
  \BibitemOpen
  \bibfield  {author} {\bibinfo {author} {\bibfnamefont {J.}~\bibnamefont
  {Bosma}}, \bibinfo {author} {\bibfnamefont {M.}~\bibnamefont {Sogaard}}, \
  and\ \bibinfo {author} {\bibfnamefont {Y.}~\bibnamefont {Zhang}},\ }\href
  {\doibase 10.1007/JHEP08(2017)051} {\bibfield  {journal} {\bibinfo  {journal}
  {JHEP}\ }\textbf {\bibinfo {volume} {08}},\ \bibinfo {pages} {051} (\bibinfo
  {year} {2017})},\ \Eprint {http://arxiv.org/abs/1704.04255} {arXiv:1704.04255
  [hep-th]} \BibitemShut {NoStop}%
\bibitem [{\citenamefont {Primo}\ and\ \citenamefont
  {Tancredi}(2017{\natexlab{b}})}]{Primo:2017ipr}%
  \BibitemOpen
  \bibfield  {author} {\bibinfo {author} {\bibfnamefont {A.}~\bibnamefont
  {Primo}}\ and\ \bibinfo {author} {\bibfnamefont {L.}~\bibnamefont
  {Tancredi}},\ }\href {\doibase 10.1016/j.nuclphysb.2017.05.018} {\bibfield
  {journal} {\bibinfo  {journal} {Nucl. Phys.}\ }\textbf {\bibinfo {volume}
  {B921}},\ \bibinfo {pages} {316} (\bibinfo {year} {2017}{\natexlab{b}})},\
  \Eprint {http://arxiv.org/abs/1704.05465} {arXiv:1704.05465 [hep-ph]}
  \BibitemShut {NoStop}%
\bibitem [{\citenamefont {Harley}\ \emph {et~al.}(2017)\citenamefont {Harley},
  \citenamefont {Moriello},\ and\ \citenamefont {Schabinger}}]{Harley:2017qut}%
  \BibitemOpen
  \bibfield  {author} {\bibinfo {author} {\bibfnamefont {M.}~\bibnamefont
  {Harley}}, \bibinfo {author} {\bibfnamefont {F.}~\bibnamefont {Moriello}}, \
  and\ \bibinfo {author} {\bibfnamefont {R.~M.}\ \bibnamefont {Schabinger}},\
  }\href {\doibase 10.1007/JHEP06(2017)049} {\bibfield  {journal} {\bibinfo
  {journal} {JHEP}\ }\textbf {\bibinfo {volume} {06}},\ \bibinfo {pages} {049}
  (\bibinfo {year} {2017})},\ \Eprint {http://arxiv.org/abs/1705.03478}
  {arXiv:1705.03478} \BibitemShut {NoStop}%
\bibitem [{\citenamefont {Lee}\ and\ \citenamefont
  {Mingulov}(2017)}]{Lee:2017ftw}%
  \BibitemOpen
  \bibfield  {author} {\bibinfo {author} {\bibfnamefont {R.~N.}\ \bibnamefont
  {Lee}}\ and\ \bibinfo {author} {\bibfnamefont {K.~T.}\ \bibnamefont
  {Mingulov}},\ }\href@noop {} {\  (\bibinfo {year} {2017})},\ \Eprint
  {http://arxiv.org/abs/1712.05173} {arXiv:1712.05173 [hep-ph]} \BibitemShut
  {NoStop}%
\bibitem [{\citenamefont {Smirnov}(2004)}]{Smirnov:2004ym}%
  \BibitemOpen
  \bibfield  {author} {\bibinfo {author} {\bibfnamefont {V.~A.}\ \bibnamefont
  {Smirnov}},\ }\href@noop {} {\bibfield  {journal} {\bibinfo  {journal}
  {Springer Tracts Mod. Phys.}\ }\textbf {\bibinfo {volume} {211}},\ \bibinfo
  {pages} {244} (\bibinfo {year} {2004})}\BibitemShut {NoStop}%
\bibitem [{\citenamefont {Smirnov}(2015)}]{Smirnov:2014hma}%
  \BibitemOpen
  \bibfield  {author} {\bibinfo {author} {\bibfnamefont {A.~V.}\ \bibnamefont
  {Smirnov}},\ }\href {\doibase 10.1016/j.cpc.2014.11.024} {\bibfield
  {journal} {\bibinfo  {journal} {Comput. Phys. Commun.}\ }\textbf {\bibinfo
  {volume} {189}},\ \bibinfo {pages} {182} (\bibinfo {year} {2015})},\ \Eprint
  {http://arxiv.org/abs/1408.2372} {arXiv:1408.2372} \BibitemShut {NoStop}%
\bibitem [{\citenamefont {Maierh{\"o}fer}\ \emph {et~al.}(2017)\citenamefont
  {Maierh{\"o}fer}, \citenamefont {Usovitsch},\ and\ \citenamefont
  {Uwer}}]{Maierhoefer:2017hyi}%
  \BibitemOpen
  \bibfield  {author} {\bibinfo {author} {\bibfnamefont {P.}~\bibnamefont
  {Maierh{\"o}fer}}, \bibinfo {author} {\bibfnamefont {J.}~\bibnamefont
  {Usovitsch}}, \ and\ \bibinfo {author} {\bibfnamefont {P.}~\bibnamefont
  {Uwer}},\ }\href {\doibase 10.1016/j.cpc.2018.04.012} {\  (\bibinfo {year}
  {2017}),\ 10.1016/j.cpc.2018.04.012},\ \Eprint
  {http://arxiv.org/abs/1705.05610} {arXiv:1705.05610} \BibitemShut {NoStop}%
\bibitem [{\citenamefont {Prausa}(2017)}]{Prausa:2017ltv}%
  \BibitemOpen
  \bibfield  {author} {\bibinfo {author} {\bibfnamefont {M.}~\bibnamefont
  {Prausa}},\ }\href {\doibase 10.1016/j.cpc.2017.05.026} {\bibfield  {journal}
  {\bibinfo  {journal} {Comput. Phys. Commun.}\ }\textbf {\bibinfo {volume}
  {219}},\ \bibinfo {pages} {361} (\bibinfo {year} {2017})},\ \Eprint
  {http://arxiv.org/abs/1701.00725} {arXiv:1701.00725} \BibitemShut {NoStop}%
\bibitem [{\citenamefont {Gituliar}\ and\ \citenamefont
  {Magerya}(2017)}]{Gituliar:2017vzm}%
  \BibitemOpen
  \bibfield  {author} {\bibinfo {author} {\bibfnamefont {O.}~\bibnamefont
  {Gituliar}}\ and\ \bibinfo {author} {\bibfnamefont {V.}~\bibnamefont
  {Magerya}},\ }\href {\doibase 10.1016/j.cpc.2017.05.004} {\bibfield
  {journal} {\bibinfo  {journal} {Comput. Phys. Commun.}\ }\textbf {\bibinfo
  {volume} {219}},\ \bibinfo {pages} {329} (\bibinfo {year} {2017})},\ \Eprint
  {http://arxiv.org/abs/1701.04269} {arXiv:1701.04269} \BibitemShut {NoStop}%
\bibitem [{\citenamefont {Meyer}(2018)}]{Meyer:2017joq}%
  \BibitemOpen
  \bibfield  {author} {\bibinfo {author} {\bibfnamefont {C.}~\bibnamefont
  {Meyer}},\ }\href {\doibase 10.1016/j.cpc.2017.09.014} {\bibfield  {journal}
  {\bibinfo  {journal} {Comput. Phys. Commun.}\ }\textbf {\bibinfo {volume}
  {222}},\ \bibinfo {pages} {295} (\bibinfo {year} {2018})},\ \Eprint
  {http://arxiv.org/abs/1705.06252} {arXiv:1705.06252} \BibitemShut {NoStop}%
\bibitem [{\citenamefont {Bern}\ \emph {et~al.}(2005)\citenamefont {Bern},
  \citenamefont {Del~Duca}, \citenamefont {Dixon},\ and\ \citenamefont
  {Kosower}}]{Bern:2004ky}%
  \BibitemOpen
  \bibfield  {author} {\bibinfo {author} {\bibfnamefont {Z.}~\bibnamefont
  {Bern}}, \bibinfo {author} {\bibfnamefont {V.}~\bibnamefont {Del~Duca}},
  \bibinfo {author} {\bibfnamefont {L.~J.}\ \bibnamefont {Dixon}}, \ and\
  \bibinfo {author} {\bibfnamefont {D.~A.}\ \bibnamefont {Kosower}},\ }\href
  {\doibase 10.1103/PhysRevD.71.045006} {\bibfield  {journal} {\bibinfo
  {journal} {Phys. Rev.}\ }\textbf {\bibinfo {volume} {D71}},\ \bibinfo {pages}
  {045006} (\bibinfo {year} {2005})},\ \Eprint
  {http://arxiv.org/abs/hep-th/0410224} {arXiv:hep-th/0410224} \BibitemShut
  {NoStop}%
\bibitem [{\citenamefont {Britto}\ \emph {et~al.}(2005)\citenamefont {Britto},
  \citenamefont {Cachazo},\ and\ \citenamefont {Feng}}]{Britto:2004nc}%
  \BibitemOpen
  \bibfield  {author} {\bibinfo {author} {\bibfnamefont {R.}~\bibnamefont
  {Britto}}, \bibinfo {author} {\bibfnamefont {F.}~\bibnamefont {Cachazo}}, \
  and\ \bibinfo {author} {\bibfnamefont {B.}~\bibnamefont {Feng}},\ }\href
  {\doibase 10.1016/j.nuclphysb.2005.07.014} {\bibfield  {journal} {\bibinfo
  {journal} {Nucl. Phys.}\ }\textbf {\bibinfo {volume} {B725}},\ \bibinfo
  {pages} {275} (\bibinfo {year} {2005})},\ \Eprint
  {http://arxiv.org/abs/hep-th/0412103} {arXiv:hep-th/0412103} \BibitemShut
  {NoStop}%
\bibitem [{\citenamefont {Cachazo}(2008)}]{Cachazo:2008vp}%
  \BibitemOpen
  \bibfield  {author} {\bibinfo {author} {\bibfnamefont {F.}~\bibnamefont
  {Cachazo}},\ }\href@noop {} {\  (\bibinfo {year} {2008})},\ \Eprint
  {http://arxiv.org/abs/0803.1988} {arXiv:0803.1988} \BibitemShut {NoStop}%
\bibitem [{\citenamefont {Panzer}(2015)}]{Panzer:2014caa}%
  \BibitemOpen
  \bibfield  {author} {\bibinfo {author} {\bibfnamefont {E.}~\bibnamefont
  {Panzer}},\ }\href {\doibase 10.1016/j.cpc.2014.10.019} {\bibfield  {journal}
  {\bibinfo  {journal} {Comput. Phys. Commun.}\ }\textbf {\bibinfo {volume}
  {188}},\ \bibinfo {pages} {148} (\bibinfo {year} {2015})},\ \Eprint
  {http://arxiv.org/abs/1403.3385} {arXiv:1403.3385} \BibitemShut {NoStop}%
\bibitem [{\citenamefont {Bernreuther}\ \emph
  {et~al.}(2005{\natexlab{a}})\citenamefont {Bernreuther}, \citenamefont
  {Bonciani}, \citenamefont {Gehrmann}, \citenamefont {Heinesch}, \citenamefont
  {Leineweber}, \citenamefont {Mastrolia},\ and\ \citenamefont
  {Remiddi}}]{Bernreuther:2004ih}%
  \BibitemOpen
  \bibfield  {author} {\bibinfo {author} {\bibfnamefont {W.}~\bibnamefont
  {Bernreuther}}, \bibinfo {author} {\bibfnamefont {R.}~\bibnamefont
  {Bonciani}}, \bibinfo {author} {\bibfnamefont {T.}~\bibnamefont {Gehrmann}},
  \bibinfo {author} {\bibfnamefont {R.}~\bibnamefont {Heinesch}}, \bibinfo
  {author} {\bibfnamefont {T.}~\bibnamefont {Leineweber}}, \bibinfo {author}
  {\bibfnamefont {P.}~\bibnamefont {Mastrolia}}, \ and\ \bibinfo {author}
  {\bibfnamefont {E.}~\bibnamefont {Remiddi}},\ }\href {\doibase
  10.1016/j.nuclphysb.2004.10.059} {\bibfield  {journal} {\bibinfo  {journal}
  {Nucl. Phys.}\ }\textbf {\bibinfo {volume} {B706}},\ \bibinfo {pages} {245}
  (\bibinfo {year} {2005}{\natexlab{a}})},\ \Eprint
  {http://arxiv.org/abs/hep-ph/0406046} {arXiv:hep-ph/0406046} \BibitemShut
  {NoStop}%
\bibitem [{\citenamefont {Bernreuther}\ \emph
  {et~al.}(2005{\natexlab{b}})\citenamefont {Bernreuther}, \citenamefont
  {Bonciani}, \citenamefont {Gehrmann}, \citenamefont {Heinesch}, \citenamefont
  {Leineweber}, \citenamefont {Mastrolia},\ and\ \citenamefont
  {Remiddi}}]{Bernreuther:2004th}%
  \BibitemOpen
  \bibfield  {author} {\bibinfo {author} {\bibfnamefont {W.}~\bibnamefont
  {Bernreuther}}, \bibinfo {author} {\bibfnamefont {R.}~\bibnamefont
  {Bonciani}}, \bibinfo {author} {\bibfnamefont {T.}~\bibnamefont {Gehrmann}},
  \bibinfo {author} {\bibfnamefont {R.}~\bibnamefont {Heinesch}}, \bibinfo
  {author} {\bibfnamefont {T.}~\bibnamefont {Leineweber}}, \bibinfo {author}
  {\bibfnamefont {P.}~\bibnamefont {Mastrolia}}, \ and\ \bibinfo {author}
  {\bibfnamefont {E.}~\bibnamefont {Remiddi}},\ }\href {\doibase
  10.1016/j.nuclphysb.2005.01.035} {\bibfield  {journal} {\bibinfo  {journal}
  {Nucl. Phys.}\ }\textbf {\bibinfo {volume} {B712}},\ \bibinfo {pages} {229}
  (\bibinfo {year} {2005}{\natexlab{b}})},\ \Eprint
  {http://arxiv.org/abs/hep-ph/0412259} {arXiv:hep-ph/0412259} \BibitemShut
  {NoStop}%
\bibitem [{\citenamefont {Bernreuther}\ \emph
  {et~al.}(2005{\natexlab{c}})\citenamefont {Bernreuther}, \citenamefont
  {Bonciani}, \citenamefont {Gehrmann}, \citenamefont {Heinesch}, \citenamefont
  {Leineweber},\ and\ \citenamefont {Remiddi}}]{Bernreuther:2005rw}%
  \BibitemOpen
  \bibfield  {author} {\bibinfo {author} {\bibfnamefont {W.}~\bibnamefont
  {Bernreuther}}, \bibinfo {author} {\bibfnamefont {R.}~\bibnamefont
  {Bonciani}}, \bibinfo {author} {\bibfnamefont {T.}~\bibnamefont {Gehrmann}},
  \bibinfo {author} {\bibfnamefont {R.}~\bibnamefont {Heinesch}}, \bibinfo
  {author} {\bibfnamefont {T.}~\bibnamefont {Leineweber}}, \ and\ \bibinfo
  {author} {\bibfnamefont {E.}~\bibnamefont {Remiddi}},\ }\href {\doibase
  10.1016/j.nuclphysb.2005.06.025} {\bibfield  {journal} {\bibinfo  {journal}
  {Nucl. Phys.}\ }\textbf {\bibinfo {volume} {B723}},\ \bibinfo {pages} {91}
  (\bibinfo {year} {2005}{\natexlab{c}})},\ \Eprint
  {http://arxiv.org/abs/hep-ph/0504190} {arXiv:hep-ph/0504190} \BibitemShut
  {NoStop}%
\bibitem [{\citenamefont {Bernreuther}\ \emph
  {et~al.}(2005{\natexlab{d}})\citenamefont {Bernreuther}, \citenamefont
  {Bonciani}, \citenamefont {Gehrmann}, \citenamefont {Heinesch}, \citenamefont
  {Mastrolia},\ and\ \citenamefont {Remiddi}}]{Bernreuther:2005gw}%
  \BibitemOpen
  \bibfield  {author} {\bibinfo {author} {\bibfnamefont {W.}~\bibnamefont
  {Bernreuther}}, \bibinfo {author} {\bibfnamefont {R.}~\bibnamefont
  {Bonciani}}, \bibinfo {author} {\bibfnamefont {T.}~\bibnamefont {Gehrmann}},
  \bibinfo {author} {\bibfnamefont {R.}~\bibnamefont {Heinesch}}, \bibinfo
  {author} {\bibfnamefont {P.}~\bibnamefont {Mastrolia}}, \ and\ \bibinfo
  {author} {\bibfnamefont {E.}~\bibnamefont {Remiddi}},\ }\href {\doibase
  10.1103/PhysRevD.72.096002} {\bibfield  {journal} {\bibinfo  {journal} {Phys.
  Rev.}\ }\textbf {\bibinfo {volume} {D72}},\ \bibinfo {pages} {096002}
  (\bibinfo {year} {2005}{\natexlab{d}})},\ \Eprint
  {http://arxiv.org/abs/hep-ph/0508254} {arXiv:hep-ph/0508254} \BibitemShut
  {NoStop}%
\bibitem [{\citenamefont {Gluza}\ \emph {et~al.}(2009)\citenamefont {Gluza},
  \citenamefont {Mitov}, \citenamefont {Moch},\ and\ \citenamefont
  {Riemann}}]{Gluza:2009yy}%
  \BibitemOpen
  \bibfield  {author} {\bibinfo {author} {\bibfnamefont {J.}~\bibnamefont
  {Gluza}}, \bibinfo {author} {\bibfnamefont {A.}~\bibnamefont {Mitov}},
  \bibinfo {author} {\bibfnamefont {S.}~\bibnamefont {Moch}}, \ and\ \bibinfo
  {author} {\bibfnamefont {T.}~\bibnamefont {Riemann}},\ }\href {\doibase
  10.1088/1126-6708/2009/07/001} {\bibfield  {journal} {\bibinfo  {journal}
  {JHEP}\ }\textbf {\bibinfo {volume} {07}},\ \bibinfo {pages} {001} (\bibinfo
  {year} {2009})},\ \Eprint {http://arxiv.org/abs/0905.1137} {arXiv:0905.1137}
  \BibitemShut {NoStop}%
\bibitem [{\citenamefont {Ablinger}\ \emph {et~al.}(2017)\citenamefont
  {Ablinger}, \citenamefont {Behring}, \citenamefont {Bl{\"u}mlein},
  \citenamefont {Falcioni}, \citenamefont {De~Freitas}, \citenamefont
  {Marquard}, \citenamefont {Rana},\ and\ \citenamefont
  {Schneider}}]{Ablinger:2017hst}%
  \BibitemOpen
  \bibfield  {author} {\bibinfo {author} {\bibfnamefont {J.}~\bibnamefont
  {Ablinger}}, \bibinfo {author} {\bibfnamefont {A.}~\bibnamefont {Behring}},
  \bibinfo {author} {\bibfnamefont {J.}~\bibnamefont {Bl{\"u}mlein}}, \bibinfo
  {author} {\bibfnamefont {G.}~\bibnamefont {Falcioni}}, \bibinfo {author}
  {\bibfnamefont {A.}~\bibnamefont {De~Freitas}}, \bibinfo {author}
  {\bibfnamefont {P.}~\bibnamefont {Marquard}}, \bibinfo {author}
  {\bibfnamefont {N.}~\bibnamefont {Rana}}, \ and\ \bibinfo {author}
  {\bibfnamefont {C.}~\bibnamefont {Schneider}},\ }\href@noop {} {\  (\bibinfo
  {year} {2017})},\ \Eprint {http://arxiv.org/abs/1712.09889}
  {arXiv:1712.09889} \BibitemShut {NoStop}%
\bibitem [{\citenamefont {Moch}\ \emph
  {et~al.}(2005{\natexlab{b}})\citenamefont {Moch}, \citenamefont
  {Vermaseren},\ and\ \citenamefont {Vogt}}]{Moch:2005tm}%
  \BibitemOpen
  \bibfield  {author} {\bibinfo {author} {\bibfnamefont {S.}~\bibnamefont
  {Moch}}, \bibinfo {author} {\bibfnamefont {J.~A.~M.}\ \bibnamefont
  {Vermaseren}}, \ and\ \bibinfo {author} {\bibfnamefont {A.}~\bibnamefont
  {Vogt}},\ }\href {\doibase 10.1016/j.physletb.2005.08.067} {\bibfield
  {journal} {\bibinfo  {journal} {Phys. Lett.}\ }\textbf {\bibinfo {volume}
  {B625}},\ \bibinfo {pages} {245} (\bibinfo {year} {2005}{\natexlab{b}})},\
  \Eprint {http://arxiv.org/abs/hep-ph/0508055} {arXiv:hep-ph/0508055}
  \BibitemShut {NoStop}%
\bibitem [{\citenamefont {Gehrmann}\ \emph {et~al.}(2006)\citenamefont
  {Gehrmann}, \citenamefont {Heinrich}, \citenamefont {Huber},\ and\
  \citenamefont {Studerus}}]{Gehrmann:2006wg}%
  \BibitemOpen
  \bibfield  {author} {\bibinfo {author} {\bibfnamefont {T.}~\bibnamefont
  {Gehrmann}}, \bibinfo {author} {\bibfnamefont {G.}~\bibnamefont {Heinrich}},
  \bibinfo {author} {\bibfnamefont {T.}~\bibnamefont {Huber}}, \ and\ \bibinfo
  {author} {\bibfnamefont {C.}~\bibnamefont {Studerus}},\ }\href {\doibase
  10.1016/j.physletb.2006.08.008} {\bibfield  {journal} {\bibinfo  {journal}
  {Phys. Lett.}\ }\textbf {\bibinfo {volume} {B640}},\ \bibinfo {pages} {252}
  (\bibinfo {year} {2006})},\ \Eprint {http://arxiv.org/abs/hep-ph/0607185}
  {arXiv:hep-ph/0607185} \BibitemShut {NoStop}%
\bibitem [{\citenamefont {Heinrich}\ \emph {et~al.}(2008)\citenamefont
  {Heinrich}, \citenamefont {Huber},\ and\ \citenamefont
  {Ma{\^i}tre}}]{Heinrich:2007at}%
  \BibitemOpen
  \bibfield  {author} {\bibinfo {author} {\bibfnamefont {G.}~\bibnamefont
  {Heinrich}}, \bibinfo {author} {\bibfnamefont {T.}~\bibnamefont {Huber}}, \
  and\ \bibinfo {author} {\bibfnamefont {D.}~\bibnamefont {Ma{\^i}tre}},\
  }\href {\doibase 10.1016/j.physletb.2008.03.028} {\bibfield  {journal}
  {\bibinfo  {journal} {Phys. Lett.}\ }\textbf {\bibinfo {volume} {B662}},\
  \bibinfo {pages} {344} (\bibinfo {year} {2008})},\ \Eprint
  {http://arxiv.org/abs/0711.3590} {arXiv:0711.3590} \BibitemShut {NoStop}%
\bibitem [{\citenamefont {Heinrich}\ \emph {et~al.}(2009)\citenamefont
  {Heinrich}, \citenamefont {Huber}, \citenamefont {Kosower},\ and\
  \citenamefont {Smirnov}}]{Heinrich:2009be}%
  \BibitemOpen
  \bibfield  {author} {\bibinfo {author} {\bibfnamefont {G.}~\bibnamefont
  {Heinrich}}, \bibinfo {author} {\bibfnamefont {T.}~\bibnamefont {Huber}},
  \bibinfo {author} {\bibfnamefont {D.~A.}\ \bibnamefont {Kosower}}, \ and\
  \bibinfo {author} {\bibfnamefont {V.~A.}\ \bibnamefont {Smirnov}},\ }\href
  {\doibase 10.1016/j.physletb.2009.06.038} {\bibfield  {journal} {\bibinfo
  {journal} {Phys. Lett.}\ }\textbf {\bibinfo {volume} {B678}},\ \bibinfo
  {pages} {359} (\bibinfo {year} {2009})},\ \Eprint
  {http://arxiv.org/abs/0902.3512} {arXiv:0902.3512} \BibitemShut {NoStop}%
\bibitem [{\citenamefont {Baikov}\ \emph {et~al.}(2009)\citenamefont {Baikov},
  \citenamefont {Chetyrkin}, \citenamefont {Smirnov}, \citenamefont {Smirnov},\
  and\ \citenamefont {Steinhauser}}]{Baikov:2009bg}%
  \BibitemOpen
  \bibfield  {author} {\bibinfo {author} {\bibfnamefont {P.~A.}\ \bibnamefont
  {Baikov}}, \bibinfo {author} {\bibfnamefont {K.~G.}\ \bibnamefont
  {Chetyrkin}}, \bibinfo {author} {\bibfnamefont {A.~V.}\ \bibnamefont
  {Smirnov}}, \bibinfo {author} {\bibfnamefont {V.~A.}\ \bibnamefont
  {Smirnov}}, \ and\ \bibinfo {author} {\bibfnamefont {M.}~\bibnamefont
  {Steinhauser}},\ }\href {\doibase 10.1103/PhysRevLett.102.212002} {\bibfield
  {journal} {\bibinfo  {journal} {Phys. Rev. Lett.}\ }\textbf {\bibinfo
  {volume} {102}},\ \bibinfo {pages} {212002} (\bibinfo {year} {2009})},\
  \Eprint {http://arxiv.org/abs/0902.3519} {arXiv:0902.3519} \BibitemShut
  {NoStop}%
\bibitem [{\citenamefont {Lee}\ \emph {et~al.}(2010)\citenamefont {Lee},
  \citenamefont {Smirnov},\ and\ \citenamefont {Smirnov}}]{Lee:2010cga}%
  \BibitemOpen
  \bibfield  {author} {\bibinfo {author} {\bibfnamefont {R.~N.}\ \bibnamefont
  {Lee}}, \bibinfo {author} {\bibfnamefont {A.~V.}\ \bibnamefont {Smirnov}}, \
  and\ \bibinfo {author} {\bibfnamefont {V.~A.}\ \bibnamefont {Smirnov}},\
  }\href {\doibase 10.1007/JHEP04(2010)020} {\bibfield  {journal} {\bibinfo
  {journal} {JHEP}\ }\textbf {\bibinfo {volume} {04}},\ \bibinfo {pages} {020}
  (\bibinfo {year} {2010})},\ \Eprint {http://arxiv.org/abs/1001.2887}
  {arXiv:1001.2887} \BibitemShut {NoStop}%
\bibitem [{\citenamefont {Gehrmann}\ \emph {et~al.}(2010)\citenamefont
  {Gehrmann}, \citenamefont {Glover}, \citenamefont {Huber}, \citenamefont
  {Ikizlerli},\ and\ \citenamefont {Studerus}}]{Gehrmann:2010ue}%
  \BibitemOpen
  \bibfield  {author} {\bibinfo {author} {\bibfnamefont {T.}~\bibnamefont
  {Gehrmann}}, \bibinfo {author} {\bibfnamefont {E.~W.~N.}\ \bibnamefont
  {Glover}}, \bibinfo {author} {\bibfnamefont {T.}~\bibnamefont {Huber}},
  \bibinfo {author} {\bibfnamefont {N.}~\bibnamefont {Ikizlerli}}, \ and\
  \bibinfo {author} {\bibfnamefont {C.}~\bibnamefont {Studerus}},\ }\href
  {\doibase 10.1007/JHEP06(2010)094} {\bibfield  {journal} {\bibinfo  {journal}
  {JHEP}\ }\textbf {\bibinfo {volume} {06}},\ \bibinfo {pages} {094} (\bibinfo
  {year} {2010})},\ \Eprint {http://arxiv.org/abs/1004.3653} {arXiv:1004.3653}
  \BibitemShut {NoStop}%
\bibitem [{\citenamefont {M.~Schabinger}()}]{RobSCETtalk}%
  \BibitemOpen
  \bibfield  {author} {\bibinfo {author} {\bibfnamefont {R.}~\bibnamefont
  {M.~Schabinger}},\ }\href@noop {} {\enquote {\bibinfo {title} {{Building
  Integrands For Massless Gauge Theory Amplitudes With Manifest Singularity
  Structure}},}\ }\bibinfo {note} {Talk presented at Soft-Collinear Effective
  Theory, March 19, 2018 at NIKHEF}\BibitemShut {NoStop}%
\bibitem [{\citenamefont {Brink}\ \emph {et~al.}(1977)\citenamefont {Brink},
  \citenamefont {Schwarz},\ and\ \citenamefont {Scherk}}]{Brink:1976bc}%
  \BibitemOpen
  \bibfield  {author} {\bibinfo {author} {\bibfnamefont {L.}~\bibnamefont
  {Brink}}, \bibinfo {author} {\bibfnamefont {J.~H.}\ \bibnamefont {Schwarz}},
  \ and\ \bibinfo {author} {\bibfnamefont {J.}~\bibnamefont {Scherk}},\ }\href
  {\doibase 10.1016/0550-3213(77)90328-5} {\bibfield  {journal} {\bibinfo
  {journal} {Nucl. Phys.}\ }\textbf {\bibinfo {volume} {B121}},\ \bibinfo
  {pages} {77} (\bibinfo {year} {1977})}\BibitemShut {NoStop}%
\bibitem [{\citenamefont {Arkani-Hamed}\ \emph {et~al.}(2015)\citenamefont
  {Arkani-Hamed}, \citenamefont {Bourjaily}, \citenamefont {Cachazo},
  \citenamefont {Postnikov},\ and\ \citenamefont
  {Trnka}}]{Arkani-Hamed:2014bca}%
  \BibitemOpen
  \bibfield  {author} {\bibinfo {author} {\bibfnamefont {N.}~\bibnamefont
  {Arkani-Hamed}}, \bibinfo {author} {\bibfnamefont {J.~L.}\ \bibnamefont
  {Bourjaily}}, \bibinfo {author} {\bibfnamefont {F.}~\bibnamefont {Cachazo}},
  \bibinfo {author} {\bibfnamefont {A.}~\bibnamefont {Postnikov}}, \ and\
  \bibinfo {author} {\bibfnamefont {J.}~\bibnamefont {Trnka}},\ }\href
  {\doibase 10.1007/JHEP06(2015)179} {\bibfield  {journal} {\bibinfo  {journal}
  {JHEP}\ }\textbf {\bibinfo {volume} {06}},\ \bibinfo {pages} {179} (\bibinfo
  {year} {2015})},\ \Eprint {http://arxiv.org/abs/1412.8475} {arXiv:1412.8475}
  \BibitemShut {NoStop}%
\bibitem [{\citenamefont {Bern}\ \emph {et~al.}(2016)\citenamefont {Bern},
  \citenamefont {Herrmann}, \citenamefont {Litsey}, \citenamefont
  {Stankowicz},\ and\ \citenamefont {Trnka}}]{Bern:2015ple}%
  \BibitemOpen
  \bibfield  {author} {\bibinfo {author} {\bibfnamefont {Z.}~\bibnamefont
  {Bern}}, \bibinfo {author} {\bibfnamefont {E.}~\bibnamefont {Herrmann}},
  \bibinfo {author} {\bibfnamefont {S.}~\bibnamefont {Litsey}}, \bibinfo
  {author} {\bibfnamefont {J.}~\bibnamefont {Stankowicz}}, \ and\ \bibinfo
  {author} {\bibfnamefont {J.}~\bibnamefont {Trnka}},\ }\href {\doibase
  10.1007/JHEP06(2016)098} {\bibfield  {journal} {\bibinfo  {journal} {JHEP}\
  }\textbf {\bibinfo {volume} {06}},\ \bibinfo {pages} {098} (\bibinfo {year}
  {2016})},\ \Eprint {http://arxiv.org/abs/1512.08591} {arXiv:1512.08591}
  \BibitemShut {NoStop}%
\bibitem [{\citenamefont {Bourjaily}\ \emph {et~al.}(2016)\citenamefont
  {Bourjaily}, \citenamefont {Franco}, \citenamefont {Galloni},\ and\
  \citenamefont {Wen}}]{Bourjaily:2016mnp}%
  \BibitemOpen
  \bibfield  {author} {\bibinfo {author} {\bibfnamefont {J.~L.}\ \bibnamefont
  {Bourjaily}}, \bibinfo {author} {\bibfnamefont {S.}~\bibnamefont {Franco}},
  \bibinfo {author} {\bibfnamefont {D.}~\bibnamefont {Galloni}}, \ and\
  \bibinfo {author} {\bibfnamefont {C.}~\bibnamefont {Wen}},\ }\href {\doibase
  10.1007/JHEP10(2016)003} {\bibfield  {journal} {\bibinfo  {journal} {JHEP}\
  }\textbf {\bibinfo {volume} {10}},\ \bibinfo {pages} {003} (\bibinfo {year}
  {2016})},\ \Eprint {http://arxiv.org/abs/1607.01781} {arXiv:1607.01781}
  \BibitemShut {NoStop}%
\bibitem [{\citenamefont {Bern}\ \emph {et~al.}(2018)\citenamefont {Bern},
  \citenamefont {Enciso}, \citenamefont {Shen},\ and\ \citenamefont
  {Zeng}}]{Bern:2018oao}%
  \BibitemOpen
  \bibfield  {author} {\bibinfo {author} {\bibfnamefont {Z.}~\bibnamefont
  {Bern}}, \bibinfo {author} {\bibfnamefont {M.}~\bibnamefont {Enciso}},
  \bibinfo {author} {\bibfnamefont {C.-H.}\ \bibnamefont {Shen}}, \ and\
  \bibinfo {author} {\bibfnamefont {M.}~\bibnamefont {Zeng}},\ }\href {\doibase
  10.1103/PhysRevLett.121.121603} {\bibfield  {journal} {\bibinfo  {journal}
  {Phys. Rev. Lett.}\ }\textbf {\bibinfo {volume} {121}},\ \bibinfo {pages}
  {121603} (\bibinfo {year} {2018})},\ \Eprint
  {http://arxiv.org/abs/1806.06509} {arXiv:1806.06509 [hep-th]} \BibitemShut
  {NoStop}%
\bibitem [{\citenamefont {Caron-Huot}\ and\ \citenamefont
  {Henn}(2014)}]{Caron-Huot:2014lda}%
  \BibitemOpen
  \bibfield  {author} {\bibinfo {author} {\bibfnamefont {S.}~\bibnamefont
  {Caron-Huot}}\ and\ \bibinfo {author} {\bibfnamefont {J.~M.}\ \bibnamefont
  {Henn}},\ }\href {\doibase 10.1007/JHEP06(2014)114} {\bibfield  {journal}
  {\bibinfo  {journal} {JHEP}\ }\textbf {\bibinfo {volume} {06}},\ \bibinfo
  {pages} {114} (\bibinfo {year} {2014})},\ \Eprint
  {http://arxiv.org/abs/1404.2922} {arXiv:1404.2922 [hep-th]} \BibitemShut
  {NoStop}%
\bibitem [{\citenamefont {Kubo}\ and\ \citenamefont
  {Milewski}(1985)}]{Kubo:1984hm}%
  \BibitemOpen
  \bibfield  {author} {\bibinfo {author} {\bibfnamefont {J.}~\bibnamefont
  {Kubo}}\ and\ \bibinfo {author} {\bibfnamefont {B.}~\bibnamefont
  {Milewski}},\ }\href {\doibase 10.1016/0550-3213(85)90224-X} {\bibfield
  {journal} {\bibinfo  {journal} {Nucl. Phys.}\ }\textbf {\bibinfo {volume}
  {B254}},\ \bibinfo {pages} {367} (\bibinfo {year} {1985})}\BibitemShut
  {NoStop}%
\bibitem [{\citenamefont {Schabinger}(2008)}]{Schabinger:2008ah}%
  \BibitemOpen
  \bibfield  {author} {\bibinfo {author} {\bibfnamefont {R.~M.}\ \bibnamefont
  {Schabinger}},\ }\href@noop {} {\  (\bibinfo {year} {2008})},\ \Eprint
  {http://arxiv.org/abs/0801.1542} {arXiv:0801.1542 [hep-th]} \BibitemShut
  {NoStop}%
\bibitem [{\citenamefont {Alday}\ \emph {et~al.}(2010)\citenamefont {Alday},
  \citenamefont {Henn}, \citenamefont {Plefka},\ and\ \citenamefont
  {Schuster}}]{Alday:2009zm}%
  \BibitemOpen
  \bibfield  {author} {\bibinfo {author} {\bibfnamefont {L.~F.}\ \bibnamefont
  {Alday}}, \bibinfo {author} {\bibfnamefont {J.~M.}\ \bibnamefont {Henn}},
  \bibinfo {author} {\bibfnamefont {J.}~\bibnamefont {Plefka}}, \ and\ \bibinfo
  {author} {\bibfnamefont {T.}~\bibnamefont {Schuster}},\ }\href {\doibase
  10.1007/JHEP01(2010)077} {\bibfield  {journal} {\bibinfo  {journal} {JHEP}\
  }\textbf {\bibinfo {volume} {01}},\ \bibinfo {pages} {077} (\bibinfo {year}
  {2010})},\ \Eprint {http://arxiv.org/abs/0908.0684} {arXiv:0908.0684
  [hep-th]} \BibitemShut {NoStop}%
\bibitem [{\citenamefont {Boels}(2010)}]{Boels:2010mj}%
  \BibitemOpen
  \bibfield  {author} {\bibinfo {author} {\bibfnamefont {R.~H.}\ \bibnamefont
  {Boels}},\ }\href {\doibase 10.1007/JHEP05(2010)046} {\bibfield  {journal}
  {\bibinfo  {journal} {JHEP}\ }\textbf {\bibinfo {volume} {05}},\ \bibinfo
  {pages} {046} (\bibinfo {year} {2010})},\ \Eprint
  {http://arxiv.org/abs/1003.2989} {arXiv:1003.2989 [hep-th]} \BibitemShut
  {NoStop}%
\bibitem [{\citenamefont {Binosi}\ and\ \citenamefont
  {Theussl}(2004)}]{Binosi:2003yf}%
  \BibitemOpen
  \bibfield  {author} {\bibinfo {author} {\bibfnamefont {D.}~\bibnamefont
  {Binosi}}\ and\ \bibinfo {author} {\bibfnamefont {L.}~\bibnamefont
  {Theussl}},\ }\href {\doibase 10.1016/j.cpc.2004.05.001} {\bibfield
  {journal} {\bibinfo  {journal} {Comput. Phys. Commun.}\ }\textbf {\bibinfo
  {volume} {161}},\ \bibinfo {pages} {76} (\bibinfo {year} {2004})},\ \Eprint
  {http://arxiv.org/abs/hep-ph/0309015} {arXiv:hep-ph/0309015} \BibitemShut
  {NoStop}%
\bibitem [{\citenamefont {Vermaseren}(1994)}]{Vermaseren:1994je}%
  \BibitemOpen
  \bibfield  {author} {\bibinfo {author} {\bibfnamefont {J.~A.~M.}\
  \bibnamefont {Vermaseren}},\ }\href {\doibase 10.1016/0010-4655(94)90034-5}
  {\bibfield  {journal} {\bibinfo  {journal} {Comput. Phys. Commun.}\ }\textbf
  {\bibinfo {volume} {83}},\ \bibinfo {pages} {45} (\bibinfo {year}
  {1994})}\BibitemShut {NoStop}%
\end{thebibliography}%

\end{document}